\documentclass[
amsmath,
prd,nofootinbib,floatfix,11pt,
]{revtex4}

\newcommand\beq{\begin{eqnarray}}
\newcommand\eeq{\end{eqnarray}}
\newcommand\missingET{E_T^{\rm miss}}
\newcommand\MET{E_T^{\rm miss}}
\newcommand\Omegahh{\Omega_{\rm DM} h^2}
\def\lsim{\mathrel{\rlap{\lower4pt\hbox{$\sim$}}
    \raise1pt\hbox{$<$}}}                
\def\gsim{\mathrel{\rlap{\lower4pt\hbox{$\sim$}}
    \raise1pt\hbox{$>$}}}            

\allowdisplaybreaks
\usepackage{graphicx}
\usepackage{setspace}

\begin{document}
\renewcommand{\theequation}{\arabic{section}.\arabic{equation}}

\title{\Large%
Non-universal gaugino masses, the 
supersymmetric little hierarchy problem, and dark matter}
\author{James E. Younkin$^1$ and Stephen P. Martin$^{1,2}$}
\affiliation{$^1$Department of Physics, Northern Illinois University, 
DeKalb IL 60115}
\affiliation{$^2$Fermi National Accelerator Laboratory, P.O. Box 500, 
Batavia IL 60510}


\begin{abstract} 
We study a class of supersymmetric models with non-universal gaugino 
masses that could arise from $F$-terms in a general combination of the 
singlet and adjoint representations of $SU(5)$. We explore models that 
satisfy present Large Hadron Collider and other bounds, showing how the 
allowed parameter space is divided into distinct ``continents". Regions 
of parameter space that ameliorate the supersymmetric little hierarchy 
problem with a small $\mu$ parameter include the usual focus point 
scenario, but also natural areas with much lighter squarks and sleptons. 
These models are continuously connected in parameter space to regions in 
which stau co-annihilation or Higgs exchange is mostly responsible for 
dark matter annihilation, and to models in which the thermal relic 
abundance is achieved by slepton-mediated annihilation, reviving the 
bulk region that is severely restricted in mSUGRA models. In hybrid or 
confluence regions, several mechanisms combine to give the requisite 
dark matter annihilation rate. In each case we study the prospects for 
direct detection of dark matter. We also comment briefly on the impact 
of recent hints for $M_h$ near 125 GeV from the LHC.
\end{abstract}

\maketitle
\tableofcontents
\baselineskip=14.9pt

\newpage
\setcounter{footnote}{1}
\setcounter{page}{1}
\setcounter{figure}{0}
\setcounter{table}{0}

\section{Introduction\label{sec:intro}}
\setcounter{footnote}{1}
\setcounter{equation}{0}

Supersymmetry \cite{primer} as an extension of the Standard Model (SM) 
provides a way of explaining why the weak scale is so small 
compared to the Planck scale or other very high energy scales in 
fundamental physics. The Large Hadron Collider (LHC) is presently engaged 
in searches 
with increasing sensitivity to supersymmetry
\cite{ATLASSUSYlimit,CMSSUSYlimit}, excluding  
significant areas of parameter space corresponding to 
lower gluino and up-squark and down-squark masses. 
Searches by underground detectors 
for a stable neutralino
lightest supersymmetric particle (LSP) as the dark matter are also 
probing interesting parts of parameter space
\cite{CDMSlimit,XENON100limit}. In addition, the 
non-observation of a Higgs boson at the CERN LEP $e^+e^-$ collider 
\cite{Schael:2006cr}
places 
a very non-trivial constraint on the supersymmetric models. These 
searches increase the lower bounds on superpartner masses, leading to an 
apparent fine-tuning problem
known as the 
supersymmetric little hierarchy 
problem. The essence of this problem is the tension between the $Z$ boson 
squared mass and the much larger soft supersymmetry-breaking squared 
masses.

In the Minimal Supersymmetric Standard Model (MSSM), this problem can be 
understood by considering the relationship between input parameters and 
the electroweak scale, which may be written as
\beq
m_Z^2 = -2(|\mu|^2 + m^2_{H_u}) + \ldots
\label{eq:m2Z}
\eeq
where the ellipses denote loop 
correction effects (which can be made small by an appropriate choice of
renormalization scale) and a tree-level contribution suppressed by $1/\tan^2\beta$. 
Here $\tan\beta$ is equal to the ratio of Higgs expectation 
values $\langle H_u \rangle/\langle H_d \rangle$, 
$\mu$ is the superpotential Higgs mass parameter, and $m_{H_u}^2$ is the soft 
supersymmetry breaking squared mass of the Higgs boson that couples to 
the top quark. The cancellation needed between $|\mu|^2$ and $m_{H_u}^2$ 
may therefore be regarded as an indication\footnote{Because there is no objective 
measure on parameter 
space, we do not attempt any detailed quantification of 
fine-tuning, but do view it as a 
qualitatively valid motivation and concern.} 
 of the fine-tuning required to obtain the observed weak scale. 

In terms of the input running gaugino masses $M_1$, $M_2$, $M_3$ 
at the apparent unification 
scale $M_U = 2 \times 10^{16}$ GeV, one finds from running two-loop 
renormalization group equations \cite{twoloopsoft} that $m_{H_u}^2$ at 
the TeV scale is approximately:
\beq
-m_{H_u}^2 &=& 1.82 M_3^2 - 0.21 M_2^2 + 0.16 M_3 M_2 
+ 0.023 M_1 M_3 + 0.006 M_1 M_2 - 0.006 M_1^2 
\nonumber
\\
&&-0.32 A_0 M_3 - 0.07 A_0 M_2 - 0.022 m_0^2.
\phantom{xxx}
\label{eq:m2Hu}
\eeq
Here we have chosen $\tan\beta = 10$ and chosen flavor-blind scalar 
squared 
masses $m_0^2$ and a universal
scalar cubic coupling parameter $A_0$ for illustration. In the 
subspace of models with unified gaugino 
masses $M_1 = M_2 = M_3 = m_{1/2}$ at $M_U$
(often 
referred to as either ``mSUGRA" or ``CMSSM" in the literature), 
lower bounds on the gluino mass imply 
a significant fine tuning in order to reconcile 
equations (\ref{eq:m2Z}) and (\ref{eq:m2Hu}). This will worsen if the LHC sets new 
limits, a plausible prospect in the very near future.

This motivates models with non-universal gaugino masses, in particular 
those in which the gluino mass parameter $M_3$ is relatively small 
compared to the wino mass parameter $M_2$. As can be seen from 
eq.~(\ref{eq:m2Hu}), a small ratio of $M_3/M_2$ can lower $-m_{H_u}^2$ 
and therefore decrease the amount of cancellation needed with $|\mu|^2$
\cite{KaneKing}. 
In this paper, we will explore some features, including dark matter 
properties, of a class of non-universal 
gaugino mass models, with particular attention to models that can 
naturally accommodate small $|\mu|$ and are therefore more attractive 
from the perspective of the supersymmetric little hierarchy problem.

Specifically, we will use a class of models that have the gauginos getting 
contributions to their masses from an $F$-term that transforms as a ${\bf 
24}$ representation of the global $SU(5)$ group that contains the SM 
gauge group \cite{SU5nonuniversal}. This yields a contribution to the 
gaugino masses in the ratio $M_1:M_2:M_3 = 1:3:-2$. The same pattern 
arises \cite{SO10nonuniversal} if the $F$-terms transform as a ${\bf 54}$ 
representation of the global $SO(10)$ group that contains $SU(5)$. In 
either case, the group may or may not be promoted to a true grand 
unification gauge group. Gaugino masses in this pattern may be added to a 
universal contribution $M_1:M_2:M_3 = 1:1:1$.

In the CMSSM, there are 
four main mechanisms that allow for dark matter annihilation that 
is efficient enough to prevent the early universe from becoming matter 
dominated too soon. First, the ``bulk" region at small $m_0$ and small 
$m_{1/2}$ allows efficient dark matter annihilation by $t$-channel 
slepton exchange. This region is under considerable pressure (if not 
eliminated entirely) by the LEP and LHC bounds, but with gaugino mass 
non-universality we will see that it is easily resurrected.
Second, there is a co-annihilation region at small 
$m_0$, in which the lightest stau co-exists and co-annihilates in the 
early universe with an LSP that is not much lighter 
\cite{staucoannihilation}. Third, a ``focus point" or ``small $\mu$" region 
\cite{focuspoint} at very large 
$m_0$, has $\mu$ small, so that mostly bino-like LSPs contain a 
subdominant but significant Higgsino admixture, allowing them to 
annihilated efficiently \cite{focuspoint} to and through weak bosons. If the
LSP mass exceeds $m_t$, then the final state of dark matter pair annihilations
can be $t\overline t$ in this case. 
Fourth, at large $\tan\beta$ the $A^0$-funnel region allows the LSPs to 
annihilate through $s$-channel pseudo-scalar Higgs exchange 
\cite{Afunnel}. In the more general MSSM, there 
is also a  possibility of LSPs that 
co-annihilate efficiently because they are wino-like 
\cite{winocoannihilation}, but this cannot be realized in the unified 
gaugino mass case. In the following, we will take the 
dark matter density to be in the conservative range
\beq
0.09 < \Omegahh < 0.13,
\label{eq:DMrange}
\eeq
where $h \approx 0.71$ is the Hubble parameter today in units of km/(sec Mpc).
It is very important that it is optional to require the predicted thermal relic 
abundance to lie in this range, because there are a variety of easy ways to 
evade this bound (see for example \cite{gelgon}).
If the predicted $\Omegahh$ for a particular model comes out lower 
than $0.09$, then
axions or some other species 
could make up the difference, and so this region of parameter space 
should not be viewed as disallowed. If $\Omegahh > 
0.13$, then one can reduce the predicted density by having the apparent LSP 
decay to a lighter singlet particle, or by invoking $R$-parity violation so that 
there is no supersymmetric dark matter at all. Still, it is interesting to 
consider the range eq.~(\ref{eq:DMrange}) as providing a minimal accommodation 
of astrophysical and cosmological observations \cite{OmegaDM}.

Modifying the universal boundary conditions of CMSSM 
can lead to interesting new conditions for dark matter; see for some recent 
examples
\cite{nonudma}-\cite{Gogoladze:2011aa} and references therein.
In \cite{compressedSUSY}, it was suggested that with a 
mixture of the two gaugino mass patterns mentioned above, chosen to 
make $\mu$ smaller, the LSP remains mostly bino-like but the observed thermal 
relic abundance of dark matter may be naturally obtained by efficient 
annihilations $\tilde N_1 \tilde N_1 \rightarrow t \overline t$ 
through $t$-channel exchange of top squarks. In these 
``compressed supersymmetry" models, the ratio of the masses of the 
heaviest and lightest superpartners is reduced
and $|\mu|$ is also 
much smaller than encountered in the CMSSM. However, 
only models continuously connected in parameter space to the CMSSM case 
were considered in \cite{compressedSUSY}.  In this paper, we will study the 
full parameter space obtained by varying over the entire 
allowed range spanned by linear combinations of the
two gaugino mass contribution patterns $M_1:M_2:M_3 = 1:1:1$ and $1:3:-2$. 
The viable parameter space is typically divided into three 
separate ``continents", with distinctive features. In the following, we will 
characterize 
these regions in terms of the size of the $\mu$ parameter, and in terms 
of the mechanisms that can allow the dark matter abundance to be in 
approximate agreement with cosmological observations.
In the larger parameter space spanned by non-zero 
$F$ terms in the ${\bf 24}$ of $SU(5)$, the dark-matter allowed regions 
mentioned above merge and are 
deformed in interesting ways. 
One of the more interesting scenarios is a parameter-space region with small 
$|\mu|$ that is 
continuously connect to the CMSSM focus-point region, but occurs at 
much lower values of $m_0$ when gaugino-mass non-universality is significant. 
There are similar regions that are not continuously connected to the CMSSM
focus point region. There are also stau co-annihilation and bulk regions,
some of which are not continuously connected to the corresponding CMSSM regions,
and which have very different prospects for dark matter direct detection.
We will 
also discuss an intriguing viable region formed by a confluence between 
the $A^0$-funnel, small-$\mu$, and stau co-annihilation regions, which 
can occupy a large chunk of parameter space, and which we refer to as the 
confluence island. However, this region is severely impacted by indirect 
constraints,
as we will see.

\section{Parameterizations and constraints\label{sec:parcon}}
\setcounter{footnote}{1}
\setcounter{equation}{0}

In this paper, we will parameterize a general combination of the universal and 
non-universal gaugino mass patterns mentioned in the Introduction in 
terms of an overall gaugino mass scale $m_{1/2}$ and an angle $\theta_{24}$, 
which we define as:
\beq
M_1 &=& m_{1/2} \left( \cos \theta_{24} + \sin \theta_{24} \right), 
\label{eq:24parametrizationa}
\\
M_2 &=& m_{1/2} \left( \cos \theta_{24} + 3 \sin \theta_{24} \right), 
\\
M_3 &=& m_{1/2} \left( \cos \theta_{24} - 2 \sin \theta_{24} \right). 
\label{eq:24parametrizationc}
\eeq
Note that $\theta_{24}=0$ corresponds to the usual unified gaugino mass 
scenario, while $\theta_{24} = \pm \pi/2$ correspond to a pure ${\bf 24}$ 
of $SU(5)$ [or ${\bf 54}$ of $SO(10)$] $F$ term contribution. Taking 
$\theta_{24} \rightarrow \theta_{24} +\pi$ flips the signs of all three 
gaugino masses simultaneously, which is physically the same as the 
original model with the signs of the 
scalar cubic couplings and $\mu$ term flipped.

Because of the high dimensionality of parameter space, it is not possible 
to do a complete study, and some choices must always be made in order to 
illustrate qualitative features. 
In this paper, we choose to include only a universal scalar squared mass
$m_0^2$, despite the fact that non-universal gaugino masses might be 
expected to be accompanied by non-universality of some kind in the scalar 
sector. We assume this scalar mass non-universality to be small, or that
it would lead to qualitative features similar to the results we obtain,
but it would certainly be interesting to consider alternatives.
In this regard, note that the scalar soft squared masses will arise 
in the form $F^*F/M_{\rm Planck}^2$, which transform in the
product of the representation $R_F$ of $F$ and its conjugate, $R_F^*$. The
direct product representation $R_F \times R_F^*$  will always include
the $SU(5)$ [or $SO(10)$] singlet representation, so that it is sensible to
take a common $m_0^2$, but it also can include other non-trivial GUT representations.
Therefore, it would be just as reasonable to explore other patterns, 
but we will not do so here.
The parameters $m_0$ and 
$\theta_{24}/\pi$ 
will be varied independently. 

In our explorations of parameter space, we used {\tt SOFTSUSY 3.1.2} 
\cite{softsusy} to generate supersymmetric spectra,\footnote{In some cases, especially at high $\tan\beta$ and large $m_0$, we found 
significant numerical differences between the model calculator we 
used, {\tt SOFTSUSY}, and an alternative, {\tt SuSpect} \cite{suspect}, 
evidently due to high sensitivity of the spectrum to the particular 
treatment of radiative corrections and translation of input parameters 
into physical masses. However, we expect the qualitative features to remain
robust.} using $m_t = 173.3$ GeV, $M_b(M_B)^{\overline{\rm MS}} = 4.25$ GeV,
and $\alpha_S(M_Z) = 0.118$. 
The program {\tt micrOMEGAs 2.2} \cite{micromegas} was used
to evaluate dark matter thermal relic abundance, 
the LSP-nucleon cross-section for dark matter detection, and implement 
constraints from other observables as follows. 
For $B \rightarrow \tau\nu$ mediated by charged Higgs bosons, 
we follow \cite{Btaunu} by taking a constraint
\beq
m_{H^+} > 13.5 \tan \beta .
\label{eq:Btaunu}
\eeq
We also consider a constraint
\beq
{\rm BR}(B_s \rightarrow \mu^+ \mu^-) < 1.1 \times 10^{-8}
\label{eq:Bsmumu}
\eeq
from \cite{Bsmumu}.
There is a well-known deviation of the anomalous magnetic moment of the 
muon measured by \cite{expgminus2}
from the Standard Model predicted values. Since we are unwilling to interpret
this measurement as ruling out the Standard Model, we follow the 
``super-conservative" attitude of \cite{Martin:2002eu}
by only requiring agreement within 5$\sigma$:
\beq
-2.5 \times 10^{-9} < \Delta a_{\mu} < 6.5 \times 10^{-9}
\label{eq:amuon}
\eeq
Here $\Delta a_\mu$ is the supersymmetric contribution to $(g_\mu -2)/2$.
In the models we consider, this only impacts the very high $\tan\beta$ case.
For $b\rightarrow s \gamma$, we implement a constraint 
\beq
2.0 \times 10^{-4} < {\rm BR}(b\rightarrow s \gamma) < 5.0 \times 10^{-4},
\label{eq:bsgamma}
\eeq 
which is also somewhat
more conservative than commonly imposed \cite{PDG}.
This is because constraints on the supersymmetric contribution to this 
observable
only apply if one accepts the additional 
assumption of minimal flavor violation in the soft supersymmetry-breaking 
sector. Because there is no reason not to expect additional sources of 
flavor violation, this constraint should not be considered mandatory. One could 
even adopt the point of view that no strict 
${\rm BR}(b\rightarrow s \gamma)$ limit applies at all, 
although relying on non-minimal 
flavor violation to counteract the minimal flavor violating contribution 
might be seen as requiring some fortuitous tuning.
In the models encountered below, $h^0$ has couplings very close to those 
of a Standard Model Higgs boson, but we use a slightly lower bound than 
the Standard Model 114.4 GeV LEP bound because of theoretical 
uncertainties in the mass 
prediction in supersymmetry; we take
\beq
m_{h^0} > 113\>\,{\rm GeV}.
\label{eq:mhlimit}
\eeq
All of the constraints mentioned above are 
implemented as contours in plots, rather than by removing the models from 
consideration, so that the impact of these indirect observables on the 
parameter space can be judged by the reader. Finally, we note that the 
CMS and ATLAS collaborations have announced significant limits 
\cite{ATLASSUSYlimit,CMSSUSYlimit} on gluino and squark 
production, although only in the context of unified gaugino mass models 
and other simplified models with large hierarchies between the gluino and LSP 
masses. Most of the models we will study evade the LHC limits by taking 
sufficiently heavy gluinos, with $M_3$ held fixed. In cases where $M_1$
is held fixed instead,  
we will also plot the contours for 
$m_{\tilde q} = 800$ and 1000 GeV for the
average squark mass on graphs below, as a very rough indication of where 
the exclusion reach of present LHC data might lie. However, because the 
superpartner mass spectrum is compressed in many of the parameter space regions 
we 
study, and the reach of the LHC can be significantly reduced in such 
cases \cite{LeCompte}, a true 
exclusion would require a dedicated study that is beyond the scope of this 
paper.

\section{Explorations with fixed $M_3$}
\setcounter{footnote}{1}
\setcounter{equation}{0}

In this section, we study
several slices of parameter space with $M_3$ held fixed at 600 GeV, so that the 
gluino mass is beyond the present reach of the LHC, in the range $1350 < M_
{\tilde g} < 1525$ depending on the masses of squarks (which contribute to the 
gluino mass in loop corrections \cite{hep-ph/9308222}).
We will consider
both moderate and large $\tan\beta$, and varying $m_0$ and $\theta_{24}$.
From eqs.~(\ref{eq:24parametrizationa})-(\ref{eq:24parametrizationc}), 
one has:
\beq
M_1 \,=\, M_3 \left( \frac{1 + \tan \theta_{24}}{1 -2 \tan \theta_{24}} 
\right) ,\qquad\quad
M_2 \,=\, M_3 \left( \frac{1 + 3 \tan \theta_{24}}{1-2 \tan \theta_{24}} 
\right) .
\eeq
Since we are holding $M_3$ fixed in this section, $|M_1|$ becomes very small 
when 
$\theta_{24}/\pi$ approaches $-1/4$ and $3/4$.
Therefore, in the graphs below we take the range for $\theta_{24}/\pi$
from $-1/4$ to $3/4$; this range 
(rather than, say, 0 to 1, or $-1/2$ to $1/2$) avoids
splitting up the continents of viable parameter space in unnatural ways. 
At 
$\theta_{24}/\pi \approx -0.102$, one has $M_2$ approaching 0, so models 
near this vertical line in figures below are excluded by LEP2 limits on 
charginos \cite{PDG}, 
providing an ocean between the two 
continents. 
At $\theta_{24}/\pi \approx 0.148$, one has $|M_1|$ and $|M_2|$ diverging
without bound, so models near this vertical line are likewise excluded, 
providing 
another gap between the two continents of viable parameter space 
on either side. 
The boundaries of the continents on either side of this line
are actually determined by the fact that small $|M_3/M_2|$ leads to small 
$|\mu|^2$, 
as can be seen from eqs.~(\ref{eq:m2Z}) and (\ref{eq:m2Hu}).

\begin{figure}[!tbp]
\includegraphics[width=0.75\textwidth]{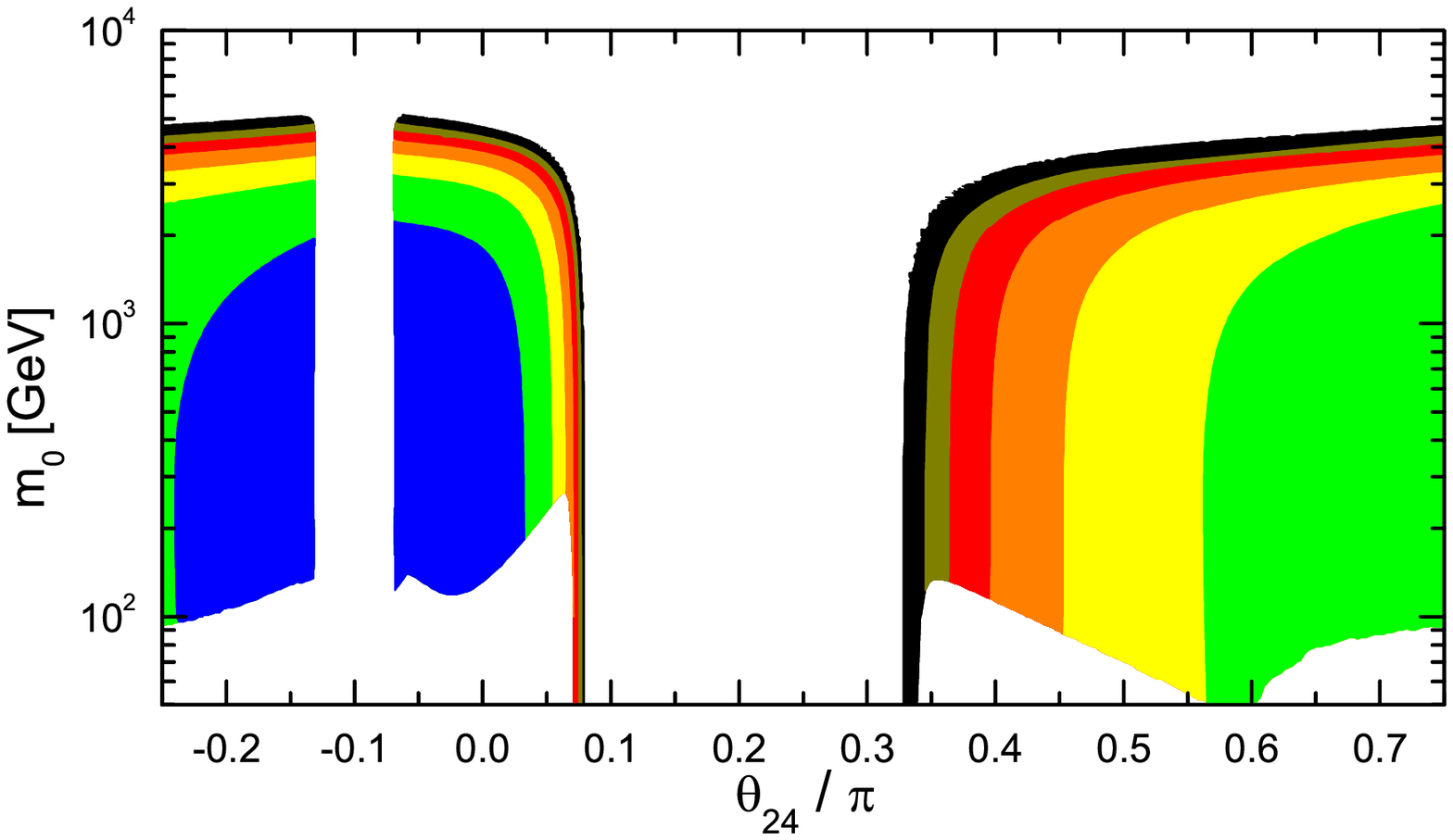} \\
\includegraphics[width=0.75\textwidth]{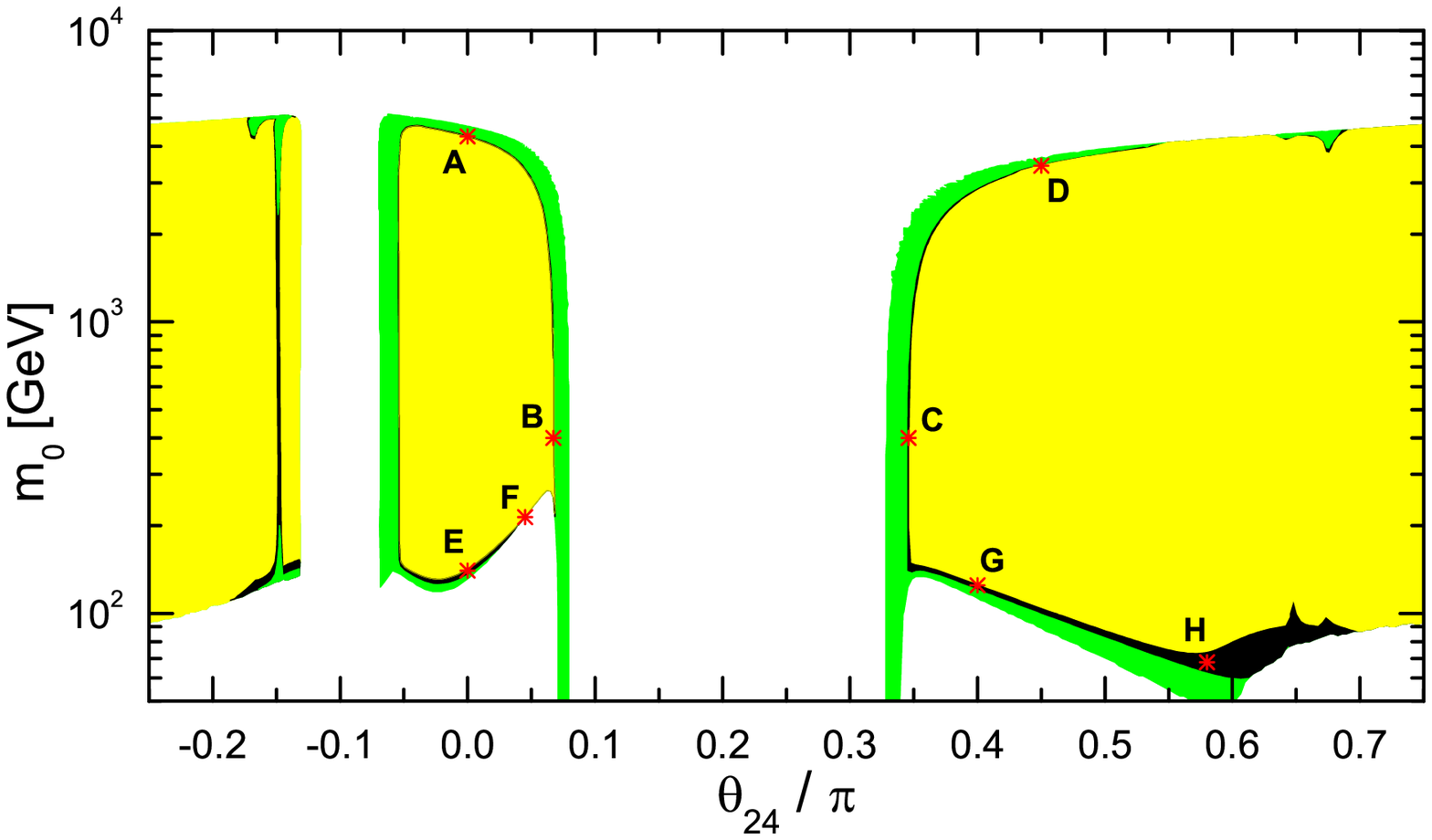}
\caption{\label{fig:60010posM3_scans}
Maps of the $\mu$ parameter (upper plot) and the predicted thermal relic 
abundance of dark matter $\Omegahh$ (lower plot) in the $\theta_{24}, 
m_0$ parameter space with fixed $M_3=-A_0 =600$ GeV, $\tan \beta = 10$, 
and $\mu>0$. In the upper plot, the dark (black) regions at the top and closest 
to the 
center represents $\mu < 300$ GeV, and successive regions lower and away from 
the 
center correspond to $300$ GeV $ < \mu < 400$ GeV (brown) and $400$ GeV $ 
< \mu < 500$ GeV (red) and so on up to the last region (blue) 
which corresponds to 800 GeV $< \mu < 900$ GeV. 
The regions left blank do not have 
viable 
electroweak symmetry breaking, do not have a neutralino as the LSP, or 
have superpartners that are too light, as described in the text. In the 
lower plot, the thin dark regions (black) correspond to the observed 
range $0.09 < \Omegahh < 0.13$. The large interior regions (yellow) 
correspond to $\Omegahh > 0.13$, while the darker shaded exterior region 
(green) has $\Omegahh < 0.09$. 
}
\end{figure}

This is illustrated in Figure \ref{fig:60010posM3_scans}, which maps the viable
regions of parameter space in the ($\theta_{24}, m_0)$ plane,
for fixed $M_3 =-A_0 =600$ GeV, $\tan \beta = 10$, 
and $\mu>0$. 
The choices of $A_0$ and $\tan\beta$ are motivated by the need to obtain a 
lightest Higgs mass above the LEP bound, and indeed all of the
indirect bounds eq.~(\ref{eq:Btaunu})-(\ref{eq:mhlimit}) are satisfied 
throughout the shaded regions shown. Regions left blank do not have viable 
electroweak symmetry breaking, do not have a neutralino as the LSP, or 
have charged superpartners (staus or charginos)
that are below the LEP2 bounds \cite{PDG}. 
The lower boundaries of the allowed continents in Figure 
\ref{fig:60010posM3_scans}, and similar figures below, 
are set by the requirement that the LSP is a 
neutralino and not a stau, as required for a supersymmetric explanation 
of the dark matter. The upper boundaries come from the requirement of proper
electroweak symmetry breaking with positive $|\mu|^2$ in the scalar potential.
On the outside parts of the plots, $-0.25 < \theta_{24}/\pi \lsim -0.16$ 
and $0.66 \lsim \theta_{24} < 0.75$, the LSP mass will be below $m_Z/2$.

The top plot in Figure \ref{fig:60010posM3_scans} is a map of the $\mu$ 
parameter, 
with lower values arguably corresponding to less fine-tuning, as described in 
the 
Introduction. Note that the MSSM corresponds to the vertical 
line $\theta_{24} = 
0$. On that line, the smallest values of $\mu$ occur at the largest allowed 
values 
of the scalar masses, near $m_0 = 4300$ GeV; this is the focus point solution. 
For smaller values of $m_0$,
the solution for electroweak symmetry breaking gives larger values of $\mu$,
but it never exceeds 900 GeV throughout the whole plot. 
The focus point region is continuously connected to
a small-$\mu$ region near $\theta_{24}/\pi = 0.65$ with a wide range of $m_0$ 
values. This region connects to a region where $\Omegahh<0.09$ which extends down to very low values of
$m_0<100$ GeV.
These models survive the requirement of a neutral (non-stau) LSP 
because here $\mu < M_1, M_2$, so that the LSP is a higgsino-like state that is much lighter than the sleptons. 

The left continent in Figure \ref{fig:60010posM3_scans} corresponds to
$M_2/M_3 < 0$, and contains small-$\mu$ regions only at large values of $m_0$, 
comparable to the CMSSM focus point. Recall that the far left of the plot 
corresponds to very light bino-like neutralino LSPs, and that the region is 
separated from the central, CMSSM-like continent by a region where $M_2$ is too 
small, leading to charginos below the LEP bound. The boundaries of that region
are very nearly vertical, because the chargino mass depends only very weakly on 
$m_0$ through loop effects.

The right continent in Figure \ref{fig:60010posM3_scans} has both $M_1/M_3$ and
$M_2/M_3$ negative, and is separated from the central CMSSM-like continent by
a region where $|M_2/M_3|$ is very large, as discussed above.
In fact, the region from $0.08 \lsim \theta_{24}/\pi \lsim 0.33$ is 
excluded because $-m_{H_u}^2$ is negative at the TeV scale, precluding 
electroweak 
symmetry breaking. The top edge of this continent is again a small-$\mu$ focus 
point-like region, and the left edge of the continent near $\theta_{24}/\pi = 
0.35$
also has small $\mu$ over a large range of $m_0$. In fact, one can see from the 
plot that the range of parameters over which $\mu$ is small is considerably 
broader
than the corresponding region on the central CMSSM-like continent, so that 
arguably this region is less fine-tuned in producing correct electroweak 
symmetry breaking.

The lower plot of Figure \ref{fig:60010posM3_scans} shows the same regions,
but now indicating whether the predicted thermal relic abundance of dark matter
is larger, smaller, or within the range of eq.~(\ref{eq:DMrange}). The region
where eq.~(\ref{eq:DMrange}) is satisfied is quite thin; this does not reflect 
any fine-tuning, but rather the fact that the dark matter relic abundance is now 
quite precisely known from experiment.
In the central continent, the interior predicts too much dark matter, due to inefficient annihilation, 
while on the boundary there 
is too little dark matter. For the 
CMSSM
at $\theta_{24} = 0$, the two allowed possibilities are the focus point region
near $m_0 = 4325$ GeV and the stau co-annihilation region near $m_0 = 140$ GeV.
The top and bottom edges of the central continent share these features. The 
vertical left edge of this continent has efficient wino co-annihilations; because of small $M_2$ the LSP has a significant wino content. The right edge 
achieves the correct dark matter abundance due to having a significant higgsino
content of the LSP because $\mu$ is small, similar to the focus point case.
Note that these regions are actually continuously connected, with hybrid features at the corners.

In the right continent  of Figure \ref{fig:60010posM3_scans}, the top and left 
edges again have small $\mu$. The bottom region is 
stau-coannihilation up to about $\theta_{\rm 24} = 0.58$, where it continuously
merges into a fatter ``bulk region", characterized not by stau coannihilations
but by slepton-mediated annihilations to leptons. Note that the 
stau-coannihilation region has $\mu$ between about 300 GeV and 700 GeV, while the 
bulk region has $\mu$ between 700 and 800 GeV. Also visible as two small peaks 
are $h^0$-resonance \cite{hresonance}
and $Z^0$ resonance regions near $\theta_{24}/\pi = 
-0.15$ and $-0.17$ respectively. The bulk region points at small $m_0$
on the right have light slepton masses not far above their LEP2 bounds.
For the left continent of Figure \ref{fig:60010posM3_scans}, the $h^0$ resonance
region cuts vertically through the entire continent, connecting to small-$\mu$ 
and bulk regions at the top and bottom of the continent, respectively.
Also shown on Figure \ref{fig:60010posM3_scans} are eight selected example models
labeled A, B, C, $\ldots$ H, for future reference.

For the models that do lie in the range of eq.~(\ref{eq:DMrange}) for 
$\Omegahh$,
it is interesting to consider the searches for direct dark matter detection,
with the most sensitive at this writing being the results from XENON100 
in ref.~\cite{XENON100limit}, which superseded earlier results from CDMS
\cite{CDMSlimit}. In Figure \ref{fig:60010posM3_scans}, we show the
ratio of the spin-independent LSP-nucleon cross section to the 
current XENON100 limit for that LSP mass, for model points in Figure 
\ref{fig:60010posM3_scans} with $\Omegahh = 0.11$.
\begin{figure}[!tbp]
\includegraphics[width=0.7\textwidth]{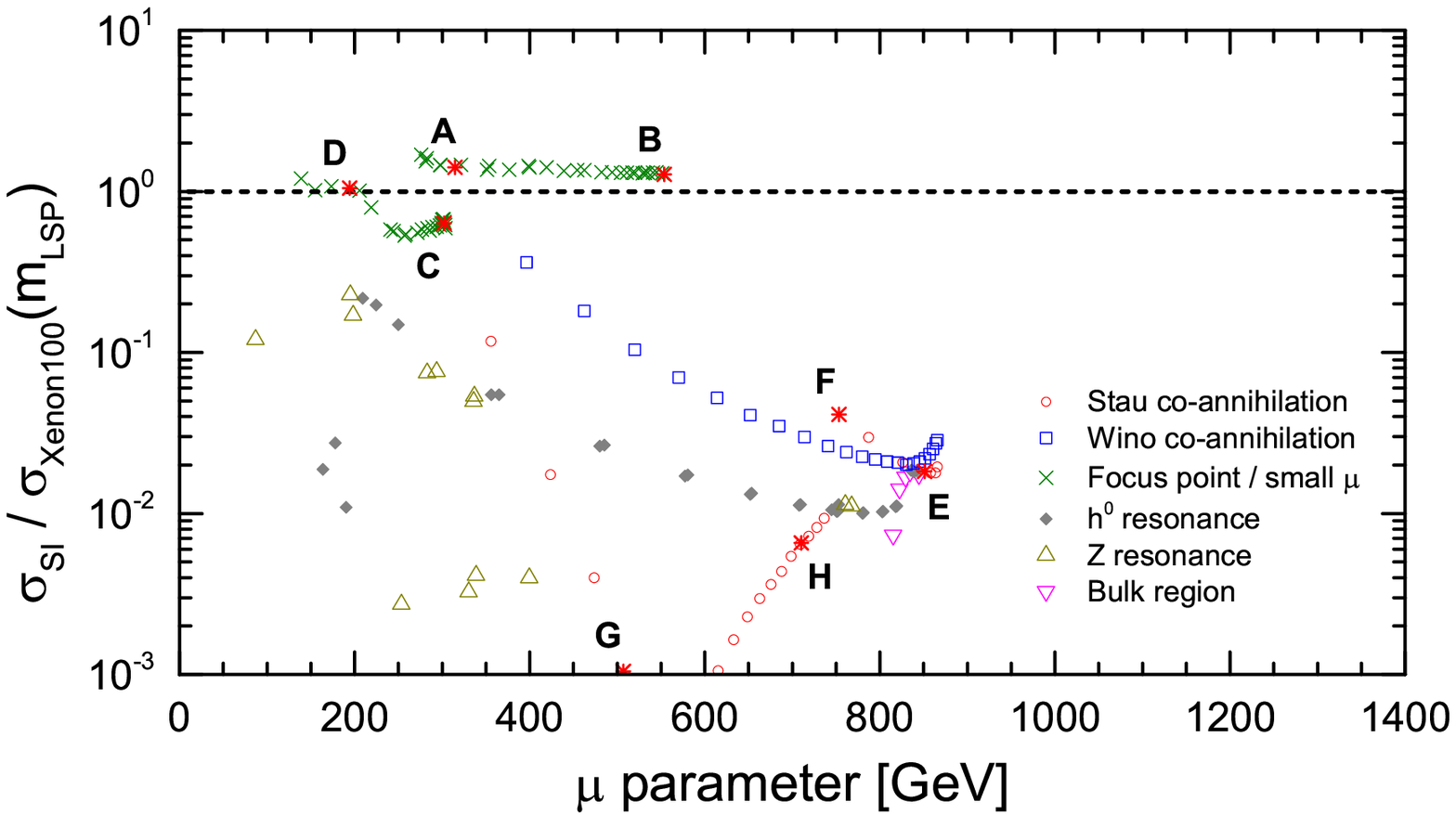} \\
\includegraphics[width=0.7\textwidth]{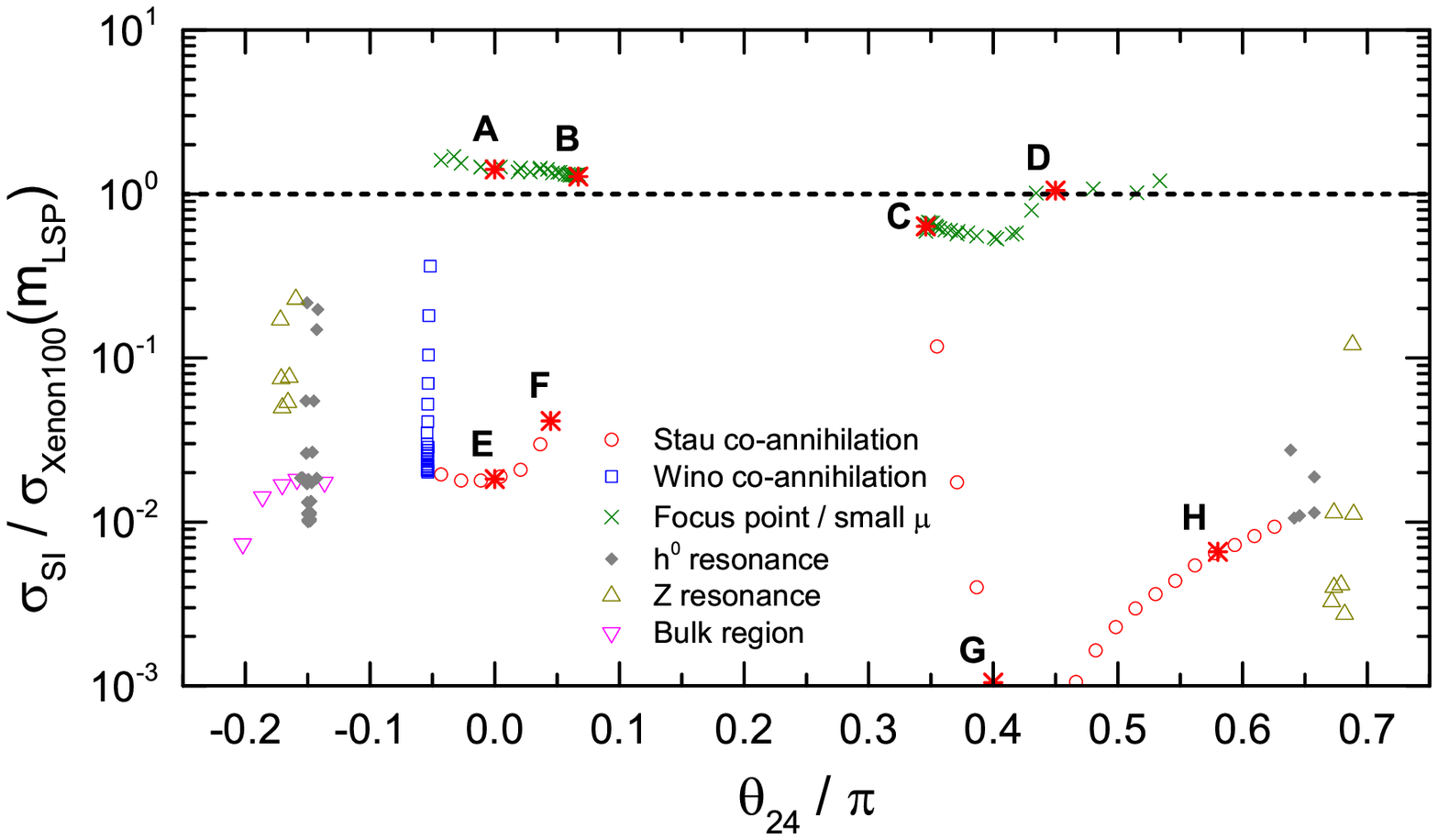}
\caption{\label{fig:60010posM3_SIcross}
The ratio of the spin-independent LSP-nucleon cross section to the 
current XENON100 limit for model points in Figure 
\ref{fig:60010posM3_scans} (with fixed $M_3=-A_0 =600$ GeV, $\tan \beta = 10$, 
and $\mu>0$ and varying $\theta_{24}, m_0$)
that have $\Omegahh = 0.11$. Different symbols are 
used for the model points according to which dark matter annihilation 
channels are most important in the early universe.
}
\end{figure}
The models are denoted by different symbols depending on what is the most 
important mechanism for dark matter annihilation in the early universe,
and plotted both as a function of $\mu$ and of $\theta_{24}$. However,
it is important to realize that the models with 
$\sigma_{SI}/\sigma_{\rm XENON100} > 1$ are not ruled out, 
for several reasons. First, the would-be LSP 
may have decayed, either to a lighter singlet or by $R$-parity 
violation,
or may have been diluted by late inflation. Second, there are significant 
uncertainties in the local density of dark matter, and especially in the
nuclear matrix elements that go into the computation of $\sigma_{SI}$.
These could easily reduce the true $\sigma_{SI}$ by more than a factor of 2.
Therefore, none of the models considered can be taken to be ruled out,
but we see that within a default interpretation, most of the small-$\mu$ models 
in the CMSSM continent, and some of those within the right continent can be said 
to be challenged by the XENON100 results. 
In contrast, the stau co-annihilation, bulk region, and wino co-annihilation 
models are not at all challenged by the present
direct dark matter searches in this model set. 
In the case of stau co-annihilation models with $\theta_{24}/\pi> 0.35$, 
$\sigma_{SI}$ does not fall monotonically 
with $\mu$, but instead falls sharply and by orders of magnitude where it 
is subject to accidental cancellations between the light and heavy Higgs 
contributions and other contributions. These cancellations are features 
of a leading order calculation, and including loop corrections and real 
emission diagrams should be expected to remove these accidental 
cancellations, but $\sigma_{SI}$ can still be so small that direct detection will be quite problematic.
The eight example model points from Figure \ref{fig:60010posM3_scans} are labeled individually. 
Note that of the small-$\mu$
models, those with $\theta_{24}/\pi \lsim 0.45$ 
most easily evade the dark matter direct detection
searches.

\begin{figure}[!tb]
\begin{minipage}[]{0.47\linewidth}
\includegraphics[width=2.9in]{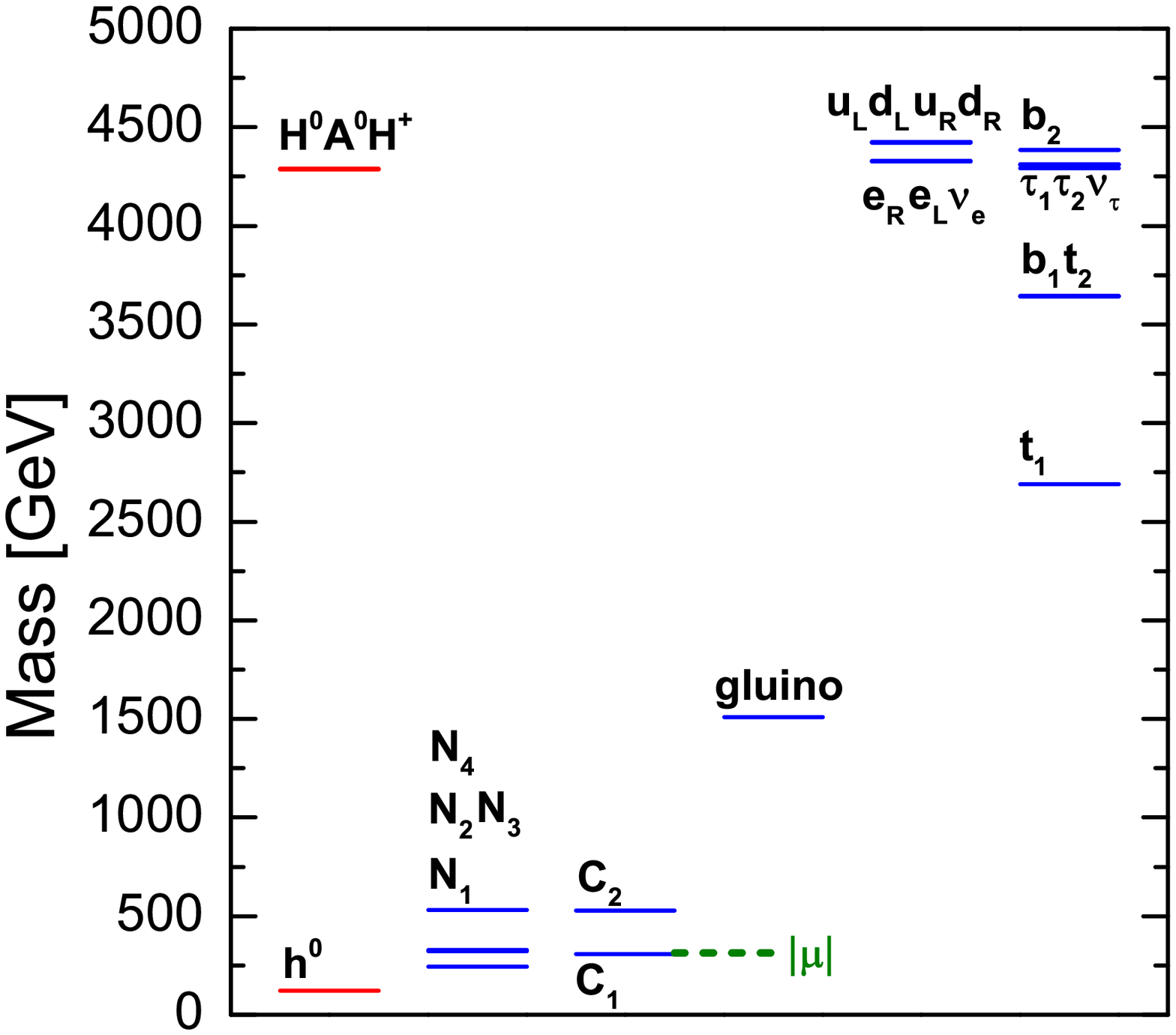}
\end{minipage}
\begin{minipage}[]{0.47\linewidth}
\includegraphics[width=2.9in]{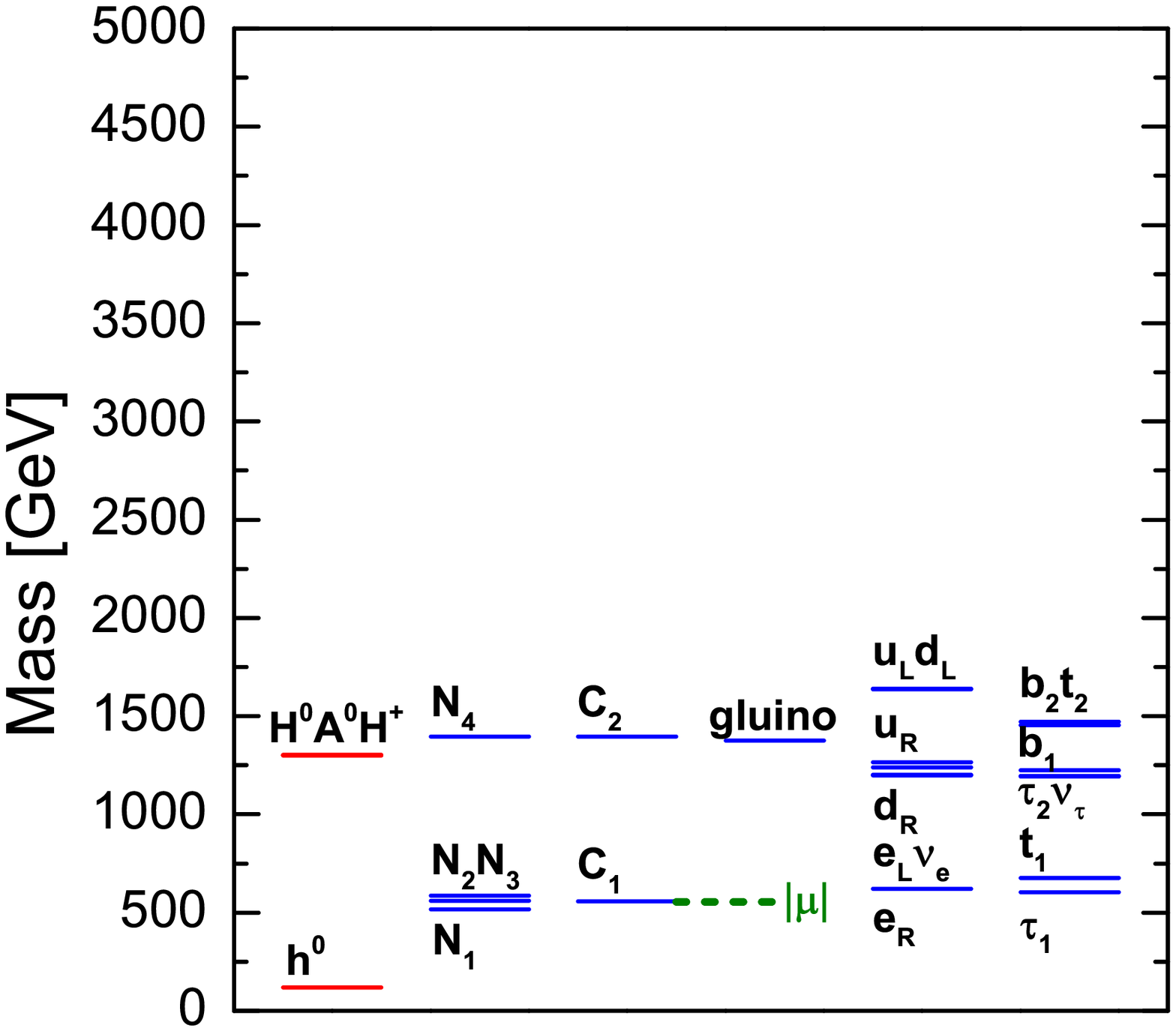}
\end{minipage}
\begin{minipage}[]{0.47\linewidth}
\includegraphics[width=2.9in]{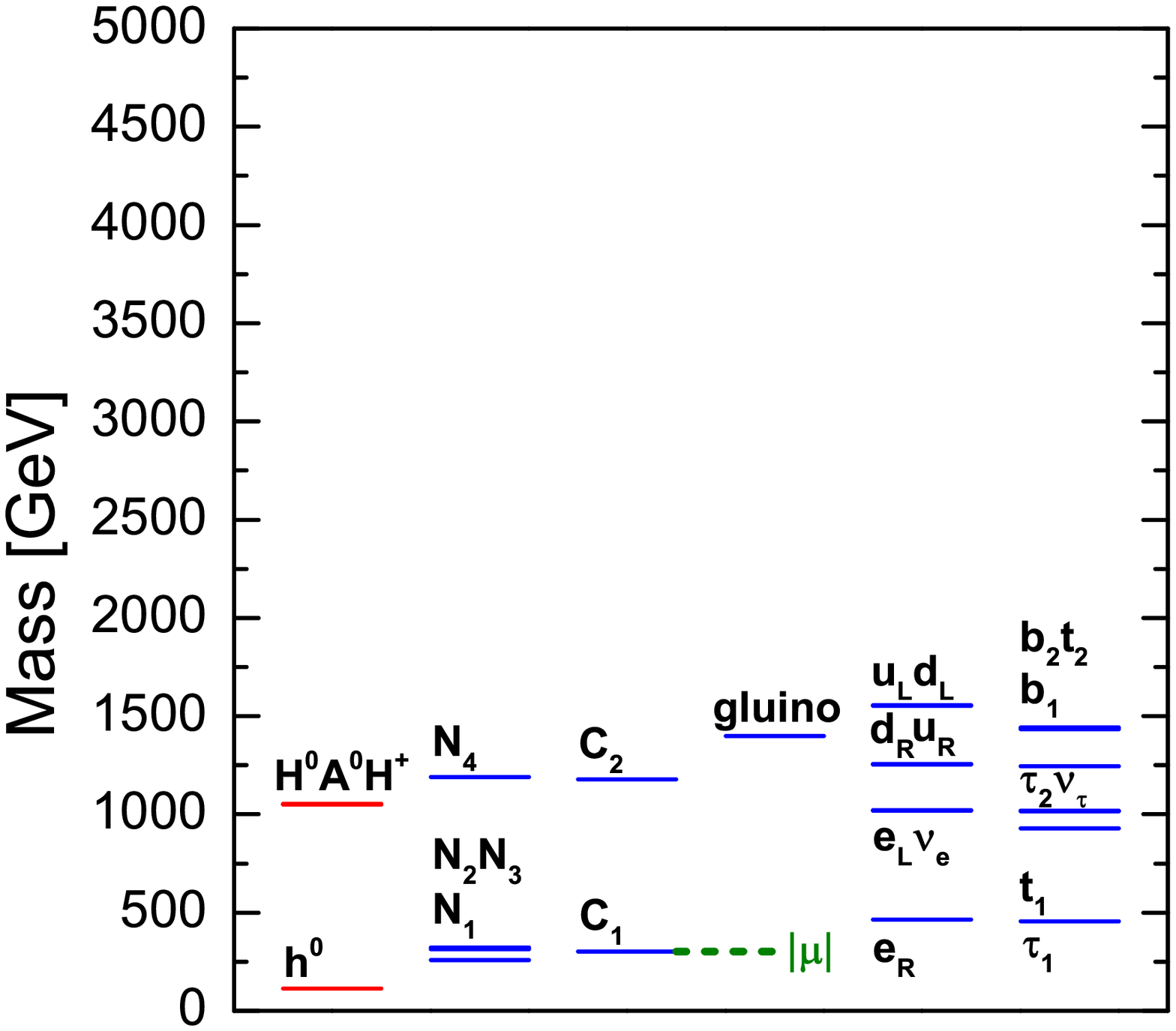}
\end{minipage}
\begin{minipage}[]{0.47\linewidth}
\includegraphics[width=2.9in]{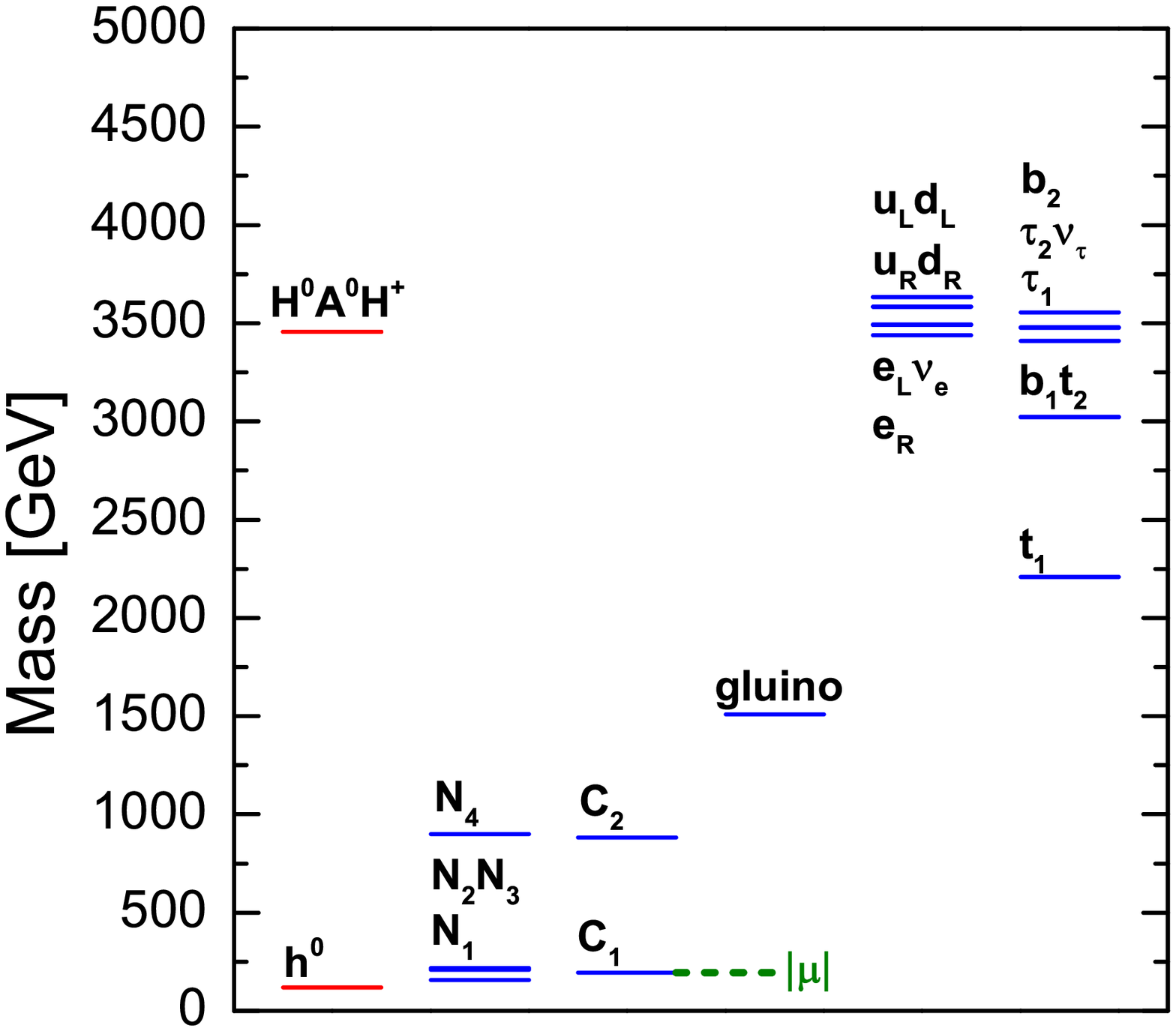}
\end{minipage}
\caption{\label{fig:60010posM3_A0pointsABCD}
Comparison of superpartner and Higgs mass spectra for 
four selected model points A,B,C,D 
(top left, top right, bottom left, and bottom right, respectively) 
in 
the small-$\mu$ dark matter regions in
Figure \ref{fig:60010posM3_scans}.}
\end{figure}

As examples, we show in Figure \ref{fig:60010posM3_A0pointsABCD}
the superpartner and Higgs mass spectra for the four selected models 
with small $\mu$ and $\Omegahh$ within the WMAP range, as labeled in 
Figures \ref{fig:60010posM3_scans} and
\ref{fig:60010posM3_SIcross}. 
The parameters for these models are 
$M_3 = -A_0 = 600$ GeV, $\tan\beta=10$, and:
\beq
\mbox{A}:&\qquad(\theta_{24}/\pi,\> m_0) \>=&(0.0,\> 4325\,{\rm 
GeV})\quad\mbox{focus point, central continent}
\\
\mbox{B}:&\qquad(\theta_{24}/\pi,\> m_0) \>=&(0.0671,\> 400\,{\rm 
GeV})\quad\mbox{small $\mu$, central continent}
\\
\mbox{C}:&\qquad(\theta_{24}/\pi,\> m_0) \>=&(0.346,\> 400\,{\rm 
GeV})\quad\mbox{small $\mu$, right continent}
\\
\mbox{D}:&\qquad(\theta_{24}/\pi,\> m_0) \>=&(0.45,\> 3440\,{\rm 
GeV})\quad\mbox{focus point, right continent}
\eeq

Model A is a CMSSM focus point model, with the well-known features of 
squarks and sleptons that are far too heavy to be seen at LHC, 
higgsino-like neutralinos and charginos that are lighter than their 
wino-like counterparts, and a Higgs boson that easily evades LEP bounds. 
Model D is qualitatively similar, although with a 
slightly lighter spectrum. This implies that in the case of model A, the 
final state of dark matter annihilation is mostly $t\overline t$, while 
it is mostly $WW$ and $ZZ$ in model D. In both of these models, discovery 
at the LHC will eventually come from gluino pair production to multi-jet 
final states. 

In contrast, models B and C both have squarks that will be accessible to 
the LHC. Both models have significant hierarchies between the right-handed 
and left-handed sleptons, and between the right-handed and left-handed 
squarks. Eventual measurements of these masses would help to confirm the 
role played by large $M_2/M_3$ in these models. Both models also have 
small mass differences between the higgsino-like neutralino and chargino 
states $\tilde N_2, \tilde N_3, \tilde C_1$ and the LSP. In the case of 
model B, co-annihilations of all of these states are important in the 
early universe, while in model B the annihilation is mostly LSP pairs to 
$t\overline t$ and $ZZ$. We also note that the spin-independent 
nucleon-LSP cross-section is a factor of 4 smaller for model C than for 
model B, and the former is below the limit from XENON100 at that mass 
while the latter is above the limit, when computed using the default 
micrOMEGAs parameters. However, another key difference between these 
models is that the lightest Higgs mass $m_h$ is in tension (near or 
possibly below) the LEP limit for model C, while it is about 119 GeV for 
model B.

\begin{figure}[!tbp]
\begin{minipage}[]{0.47\linewidth}
\includegraphics[width=2.9in]{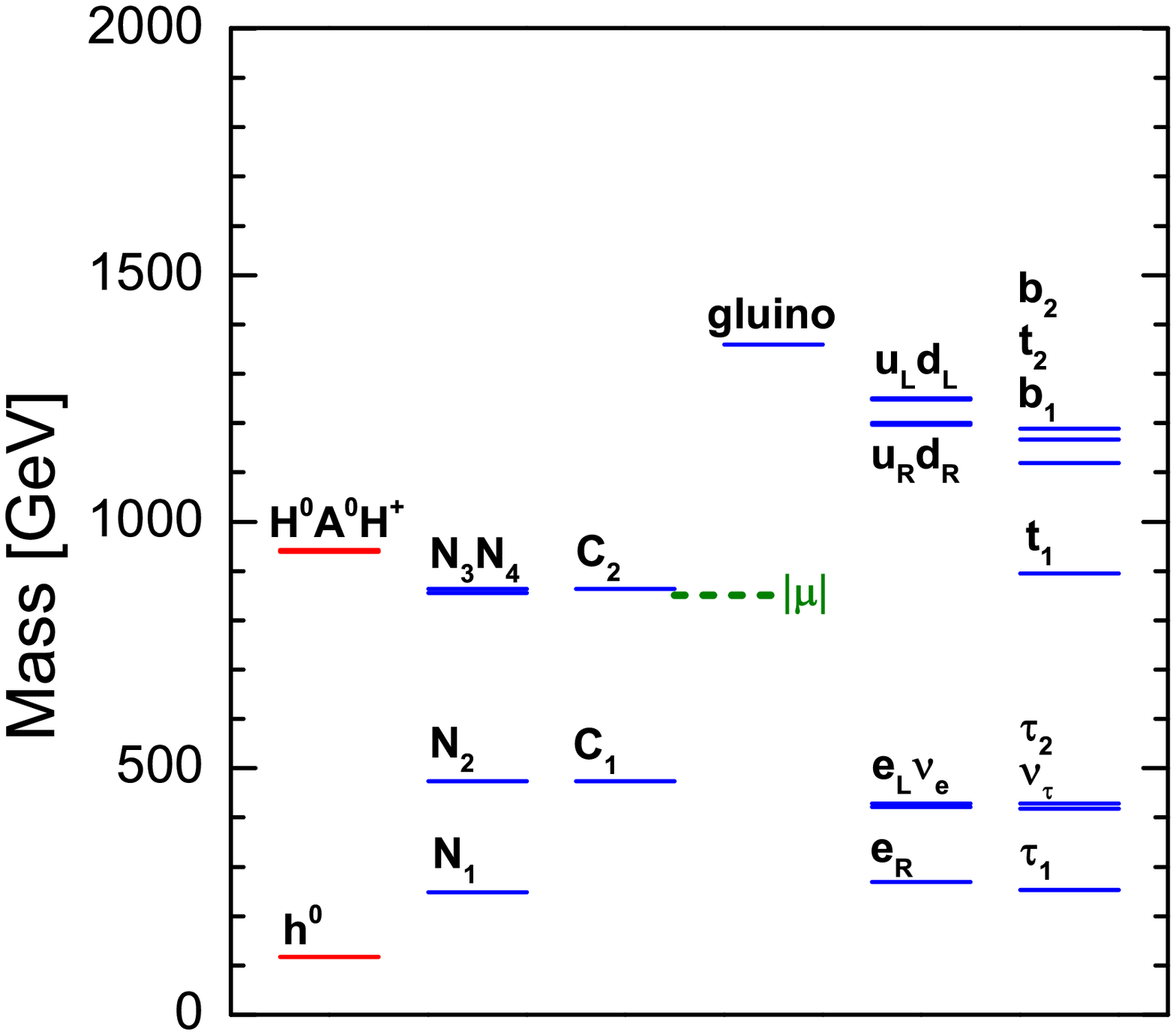}
\end{minipage}
\begin{minipage}[]{0.47\linewidth}
\includegraphics[width=2.9in]{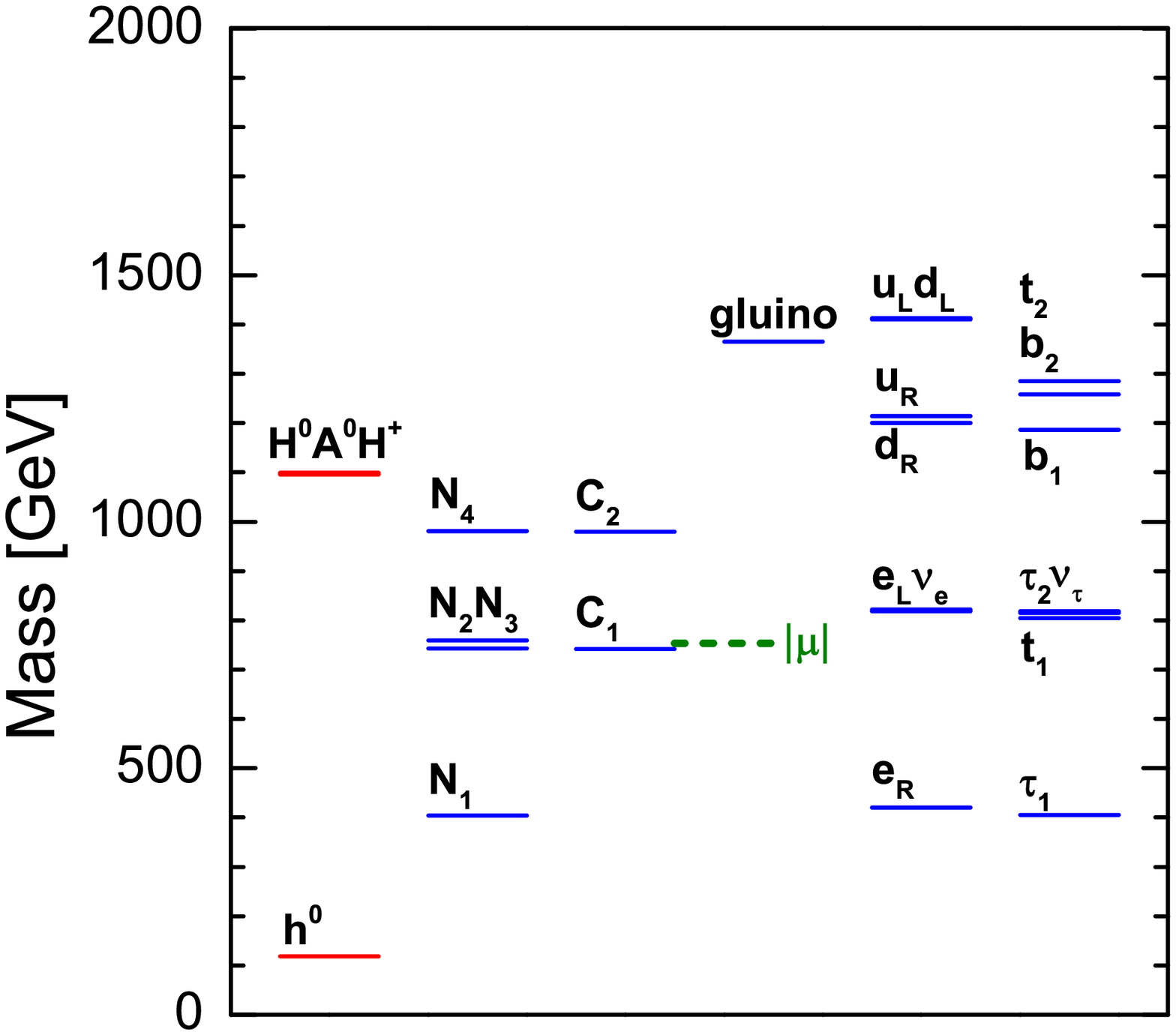}
\end{minipage}
\begin{minipage}[]{0.47\linewidth}
\includegraphics[width=2.9in]{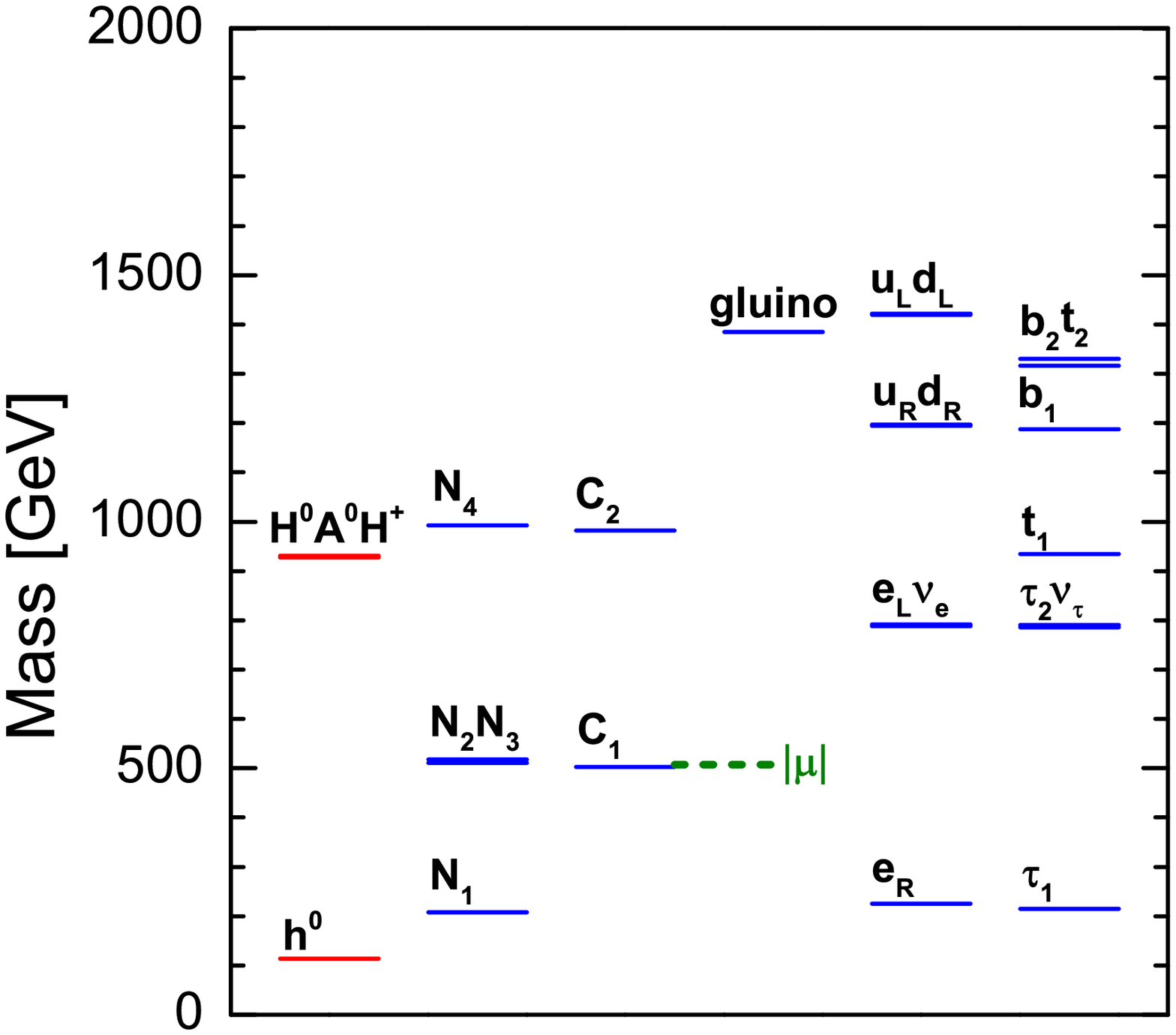}
\end{minipage}
\begin{minipage}[]{0.47\linewidth}
\includegraphics[width=2.9in]{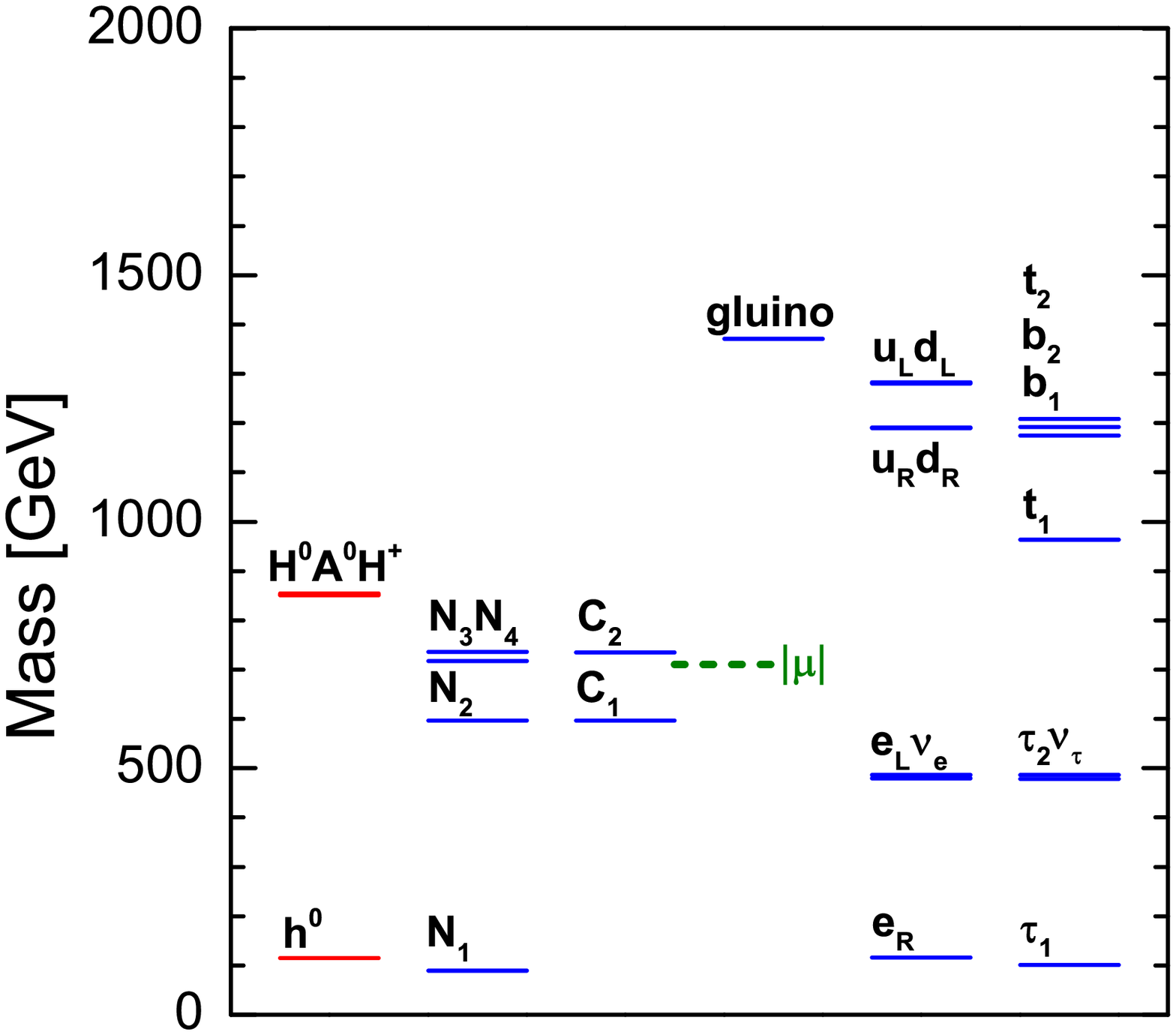}
\end{minipage}
\caption{\label{fig:60010posM3_A0pointsEFGH}
Comparison of superpartner and Higgs mass spectra for 
four selected model points E,F,G,H 
(top left, top right, bottom left, and bottom right, respectively) 
in the stau co-annihilation 
and bulk (slepton-mediated) dark matter regions in
Figure \ref{fig:60010posM3_scans}.}
\end{figure}

We also show in Figure \ref{fig:60010posM3_A0pointsEFGH} the superpartner 
and 
Higgs mass spectra for the four selected models E,F,G,H with small $m_0$, 
leading 
to stau co-annihilation and slepton-mediated dark matter regions
within the $\Omegahh$ range from WMAP.
The parameters for these models are 
$M_3 = -A_0 = 600$ GeV, $\tan\beta=10$, and:
\beq
\mbox{E}:&\qquad(\theta_{24}/\pi,\> m_0) \>=&(0.0,\> 140\,{\rm 
GeV})\quad\mbox{stau co-annihilation, central continent}
\\
\mbox{F}:&\qquad(\theta_{24}/\pi,\> m_0) \>=&(0.045,\> 215\,{\rm 
GeV})\quad\mbox{stau co-annihilation, central continent}
\\
\mbox{G}:&\qquad(\theta_{24}/\pi,\> m_0) \>=&(0.4,\> 125\,{\rm 
GeV})\quad\mbox{stau co-annihilation, right continent}
\\
\mbox{H}:&\qquad(\theta_{24}/\pi,\> m_0) \>=&(0.58,\> 68\,{\rm 
GeV})\quad\mbox{bulk (slepton-mediated), right continent}
\eeq
Model E is a typical stau co-annihilation scenario within mSUGRA,
with a small mass difference $m_{\tilde \tau_1} - m_{\tilde N_1} = 4.7$ 
GeV. It is beyond the reach of present LHC data, but will be eventually
discovered in the jets+$\MET$ searches. The spin-independent nucleon-LSP
cross-section is about a factor of 50 smaller than what would be 
necessary to probe this model with the published XENON100 searches.
Model F is also a stau co-annihilation model, but with a much more 
compressed spectrum. In this model $m_{\tilde \tau_1} - m_{\tilde N_1}$ 
is less than 
1 GeV. This small mass difference is needed in order for the 
co-annihilations to be efficient, given the larger LSP mass,
and it means that the lighter stau will have only four-body decays
$\tilde \tau_1 \rightarrow \nu_\tau \overline \nu_\ell \ell \tilde N_1$
and will be stable on time scales relevant for colliders. However, direct 
production of stau pairs is very small, and
staus occur only rarely in gluino and squark decays in this model, so the 
initial discover at LHC will again come from jets+$\MET$ searches. The 
spin-independent nucleon-LSP cross-section is only a factor of 8 below
the present XENON100 limit for model F. Both models E and F have $m_h$ 
well above the LEP limit.

Model G, in contrast, has $\sigma_{SI}$ about three orders of magnitude 
below the XENON100 limit, so dark matter direct detection will probably 
be impossible in the foreseeable future. Otherwise, it is 
similar to the CMSSM model E, with the main qualitative differences being 
a Higgs 
mass that is closer to the LEP limit of 114 GeV and a somewhat larger 
mass difference $m_{\tilde \tau_1} - m_{\tilde N_1} = 7.0$ GeV.
Model H is a bulk region model, with slepton-mediated annihilations
$\tilde N_1 \tilde N_1 \rightarrow e^+e^-$, $\mu^+\mu^-$, and $\tau^+ 
\tau^-$ mostly responsible for dark matter annihilation in the early 
universe. The lightest stau is at $m_{\tilde \tau_1} = 102$ GeV,
and about 13 GeV heavier than the LSP. Despite the low LSP mass, this 
model has $\sigma_{SI}$ more than 2 orders of magnitude below the present 
XENON100 bounds. The LHC signatures of models G and H should be very 
similar to model E, and a careful measurement of the slepton-LSP mass 
differences will be crucial for inferring their dark matter properties.

\clearpage

\begin{figure}[!tbp]
\includegraphics[width=0.7\textwidth]{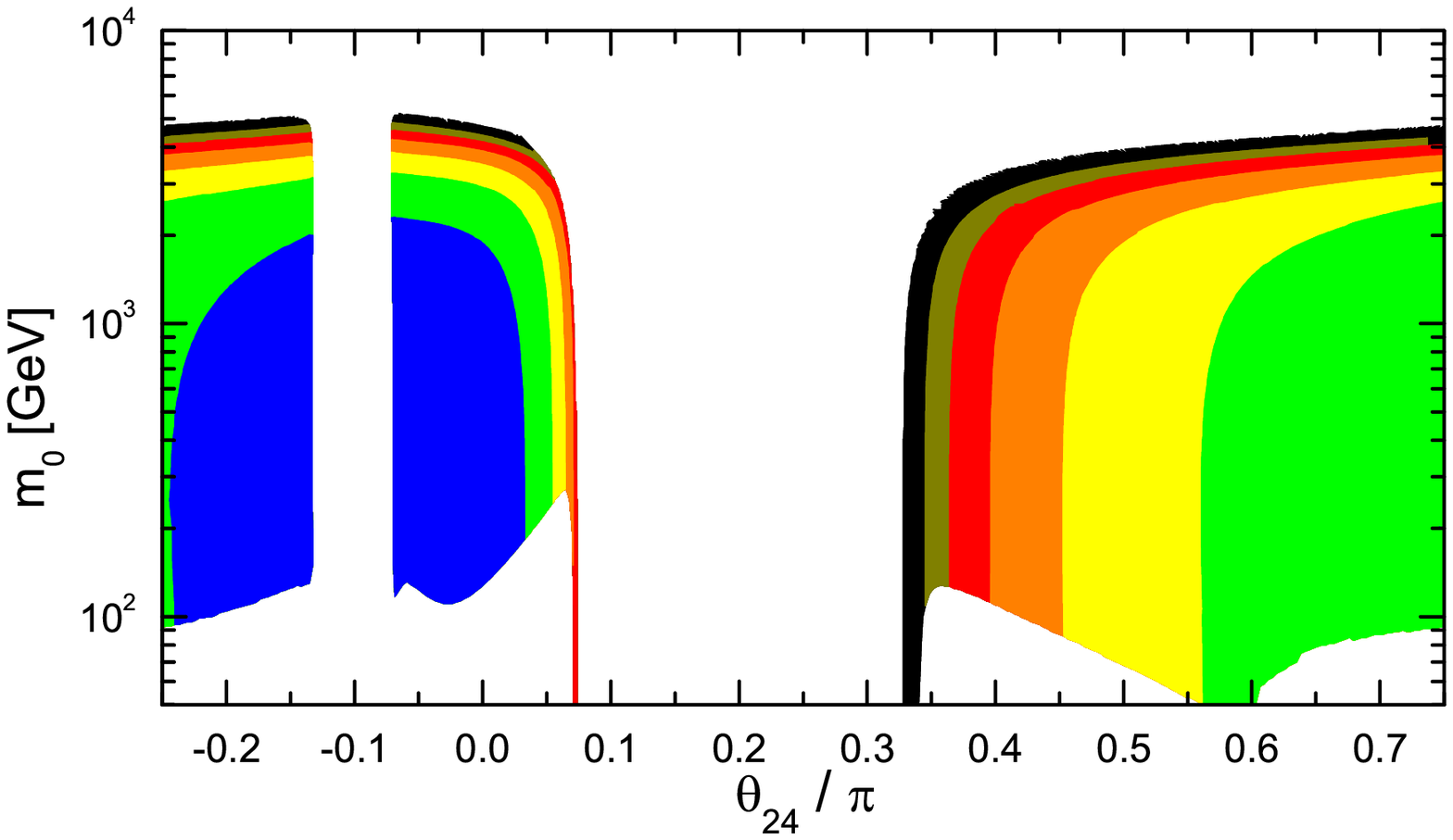} \\
\includegraphics[width=0.7\textwidth]{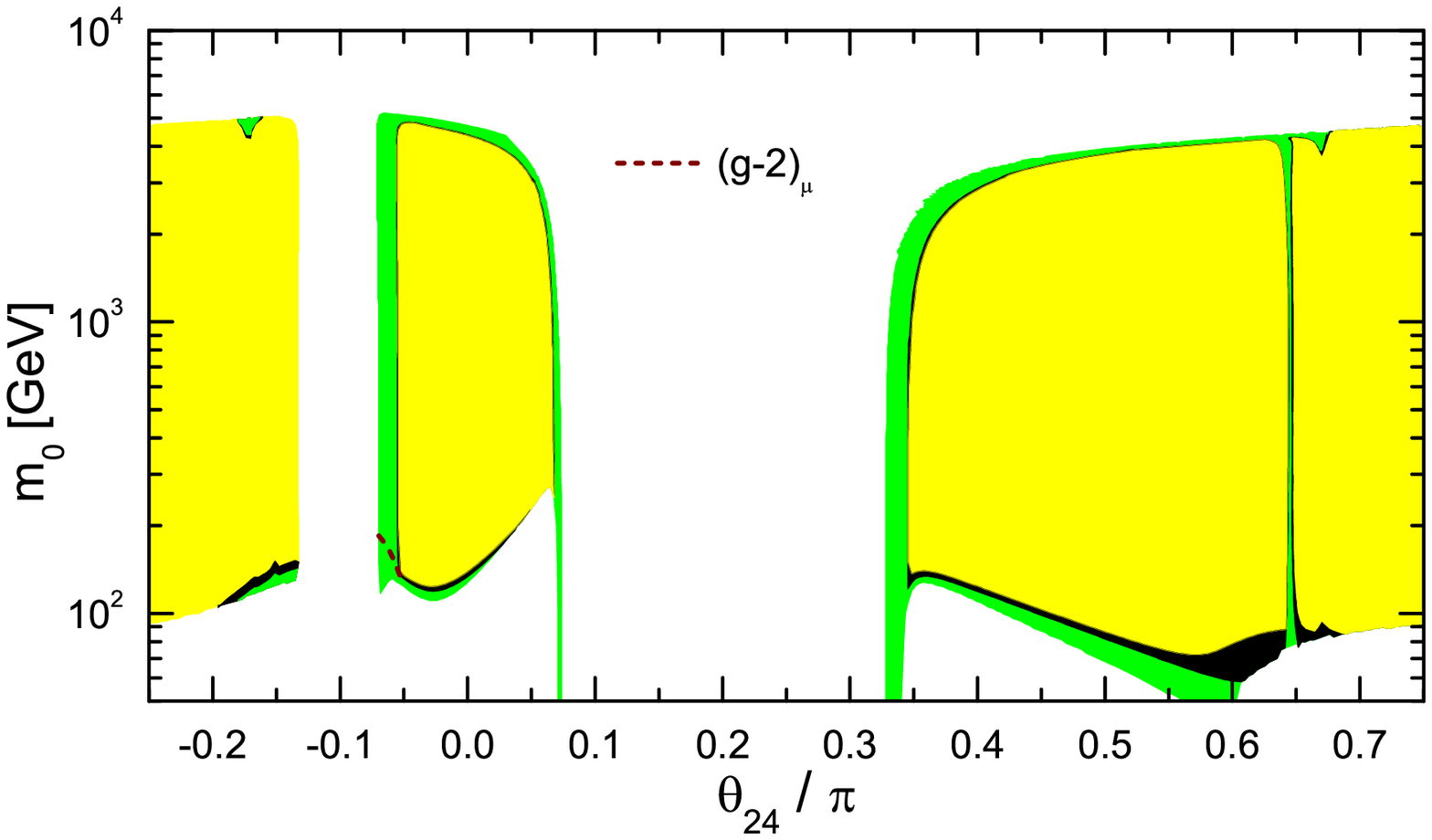}
\caption{\label{fig:60010negM3_scans}
Maps of the $\mu$ parameter (upper plot) and the predicted thermal relic 
abundance of dark matter $\Omegahh$ (lower plot) in the $\theta_{24}, 
m_0$ parameter space with fixed $M_3=-A_0 =600$ GeV, $\tan \beta = 10$, 
and $\mu<0$.  The correspondence of shaded regions to $|\mu|$ in the 
upper plot and to $\Omega_{\rm DM} h^2$ in the lower plot are the same as in
Figure \ref{fig:60010posM3_scans}. The small corner near 
$\theta_{24}/\pi = -0.07$ and $m_0 = 150$ GeV is disfavored by the 
anomalous magnetic moment of the muon.}
\end{figure}

It is also interesting to consider the same slicing through parameter space
but with $\mu<0$. The map of $|\mu|$ and $\Omegahh$ for this case is shown in
Figure \ref{fig:60010negM3_scans}.
The qualitative features for this case are very similar to the positive
$\mu$ case. In a small corner of the allowed central continent for $m_0 < 200$
GeV and $\theta_{24}/\pi$ near $-0.07$, the anomalous
magnetic moment of the muon constraint eliminates some models; 
the other indirect 
constraints from eqs.~(\ref{eq:Btaunu})-(\ref{eq:mhlimit}) do not play a 
role for these models. 
Note also
that in Figure \ref{fig:60010negM3_scans}, the $h^0$ resonance region now
is more pronounced on the right continent 
(rather than the left as for $\mu>0$), providing 
solutions with $\Omegahh < 0.13$ near $\theta_{24}/\pi = 0.64$ that are 
continuously connected to the bulk region.

Figure \ref{fig:60010posM3_scans} shows the
ratio of the spin-independent LSP-nucleon cross section to the 
current XENON100 limit for that LSP mass, for model points in Figure 
\ref{fig:60010posM3_scans} with $\Omegahh = 0.11$.
\begin{figure}[!tbp]
\includegraphics[width=0.7\textwidth]{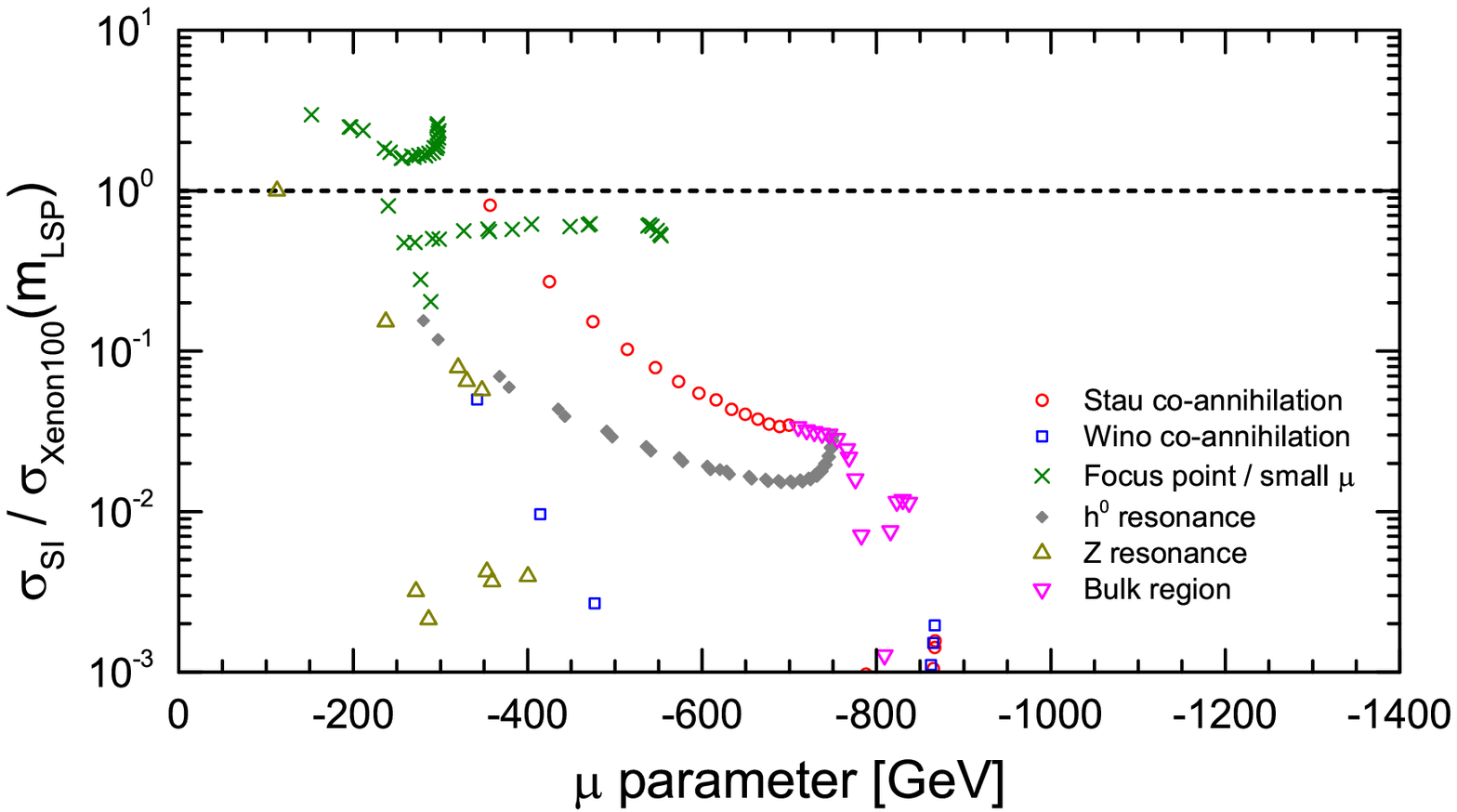} \\
\includegraphics[width=0.7\textwidth]{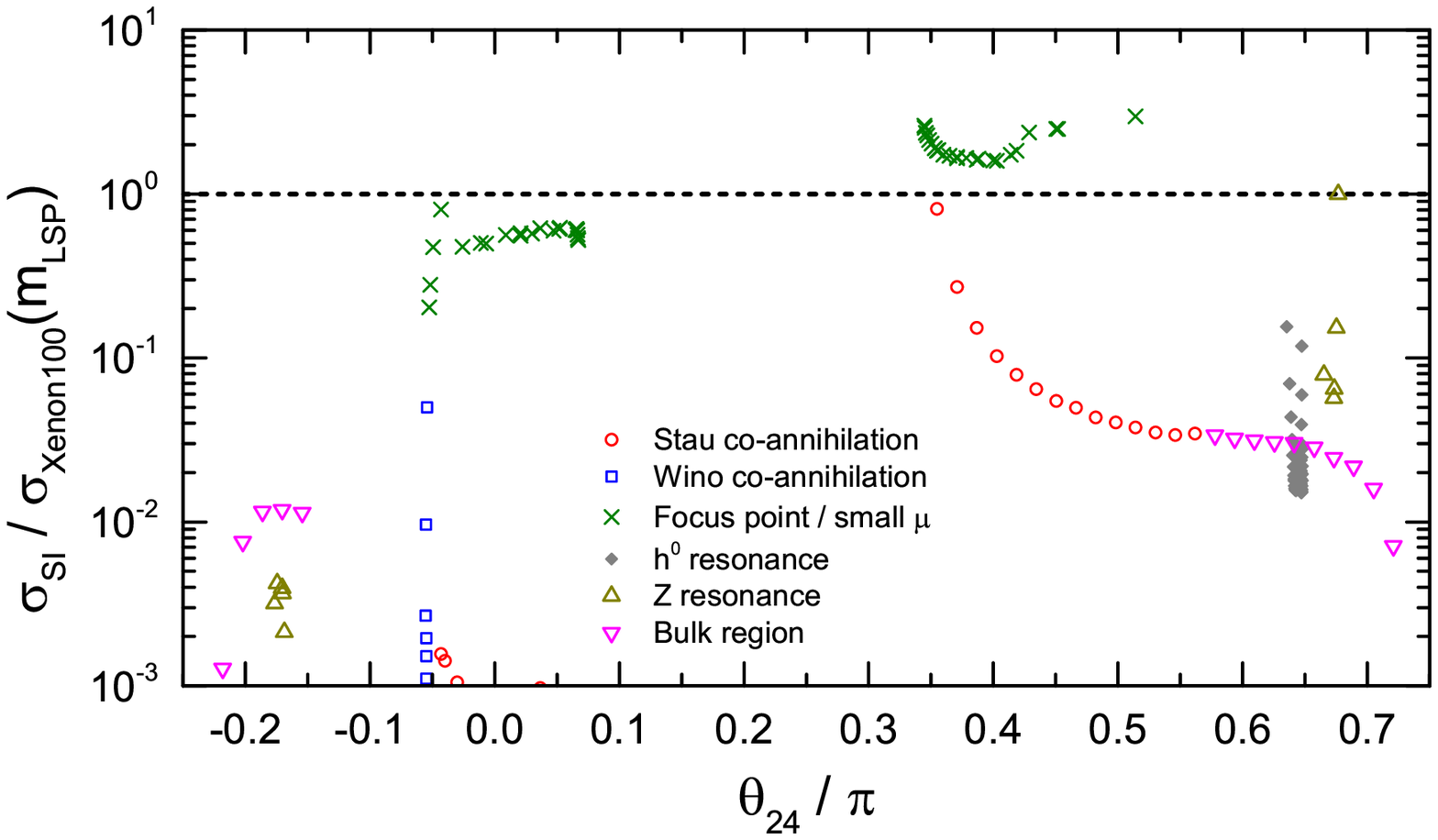}
\caption{\label{fig:60010negM3_SIcross}
The ratio of the spin-independent LSP-nucleon cross section to the 
current XENON100 limit for the model points in Figure 
\ref{fig:60010negM3_scans} (with fixed $M_3=-A_0 =600$ GeV, $\tan \beta = 10$, 
and $\mu<0$ and varying $\theta_{24}, 
m_0$)  that have $\Omegahh = 0.11$. Different symbols are 
used for the model points according to which dark matter annihilation 
channels are most important in the early universe.
}
\end{figure}
Here the most important feature that is different from the $\mu>0$ case is that
the small-$|\mu|$ models that are least susceptible to direct detection at
XENON100 for $\mu<0$ 
are those on the main, CMSSM-like continent. The focus 
point/small-$\mu$ models on the right, large-$\theta_{24}$ continent are 
nominally
above the XENON100 limit, although we reiterate that this cannot be considered
an exclusion.

\begin{figure}[!tbp]
\includegraphics[width=0.7\textwidth]{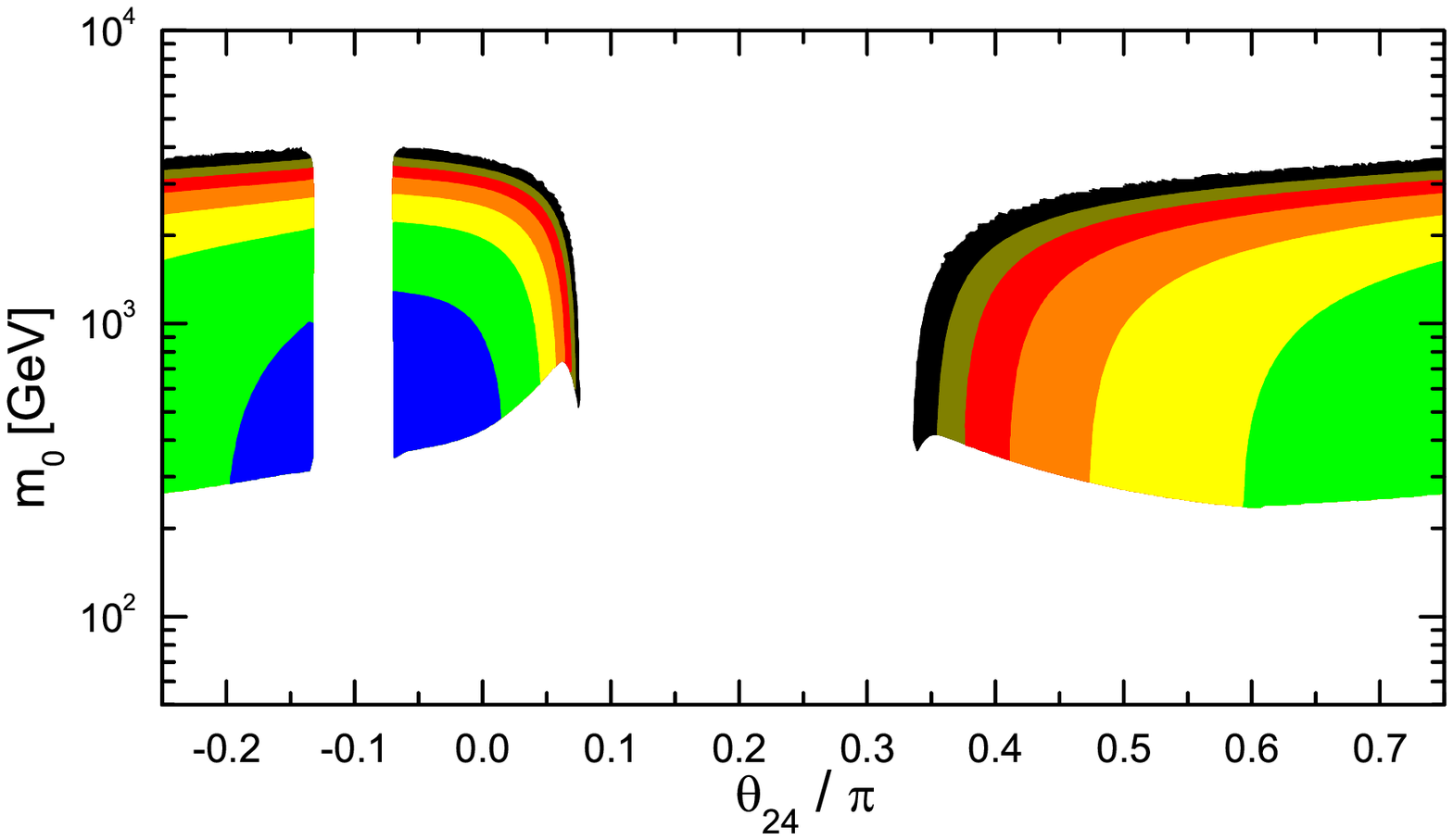} \\
\includegraphics[width=0.7\textwidth]{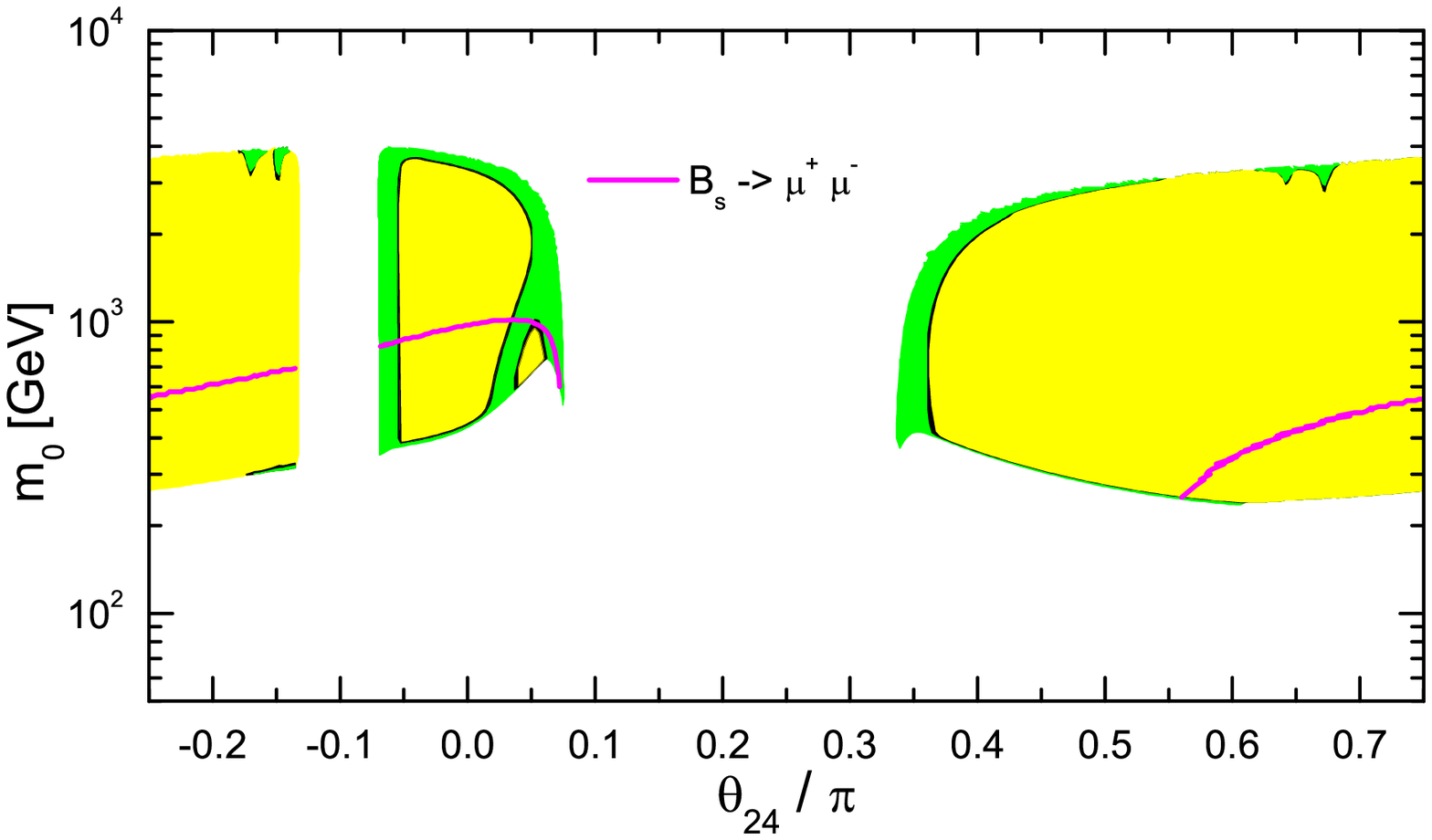}
\caption{\label{fig:60045posM3_scans}
Maps of the $\mu$ parameter (upper plot) and the predicted thermal relic 
abundance of dark matter $\Omegahh$ (lower plot) in the $\theta_{24}, 
m_0$ parameter space with fixed $M_3=-A_0 =600$ GeV, $\tan \beta = 45$, 
and $\mu>0$. 
The correspondence of shaded regions to $\mu$ in the 
upper plot and to $\Omega_{\rm DM} h^2$ in the lower plot are the same as in
Figures \ref{fig:60010posM3_scans} and \ref{fig:60010negM3_scans}. 
The regions under the contour in the lower plot are disfavored by
the limit on BR$(B_s\rightarrow \mu^+\mu^-)$.
}
\end{figure}

We now consider the impact of large $\tan\beta$ on the 
allowed parameter space. To illustrate this, we consider in 
Figure \ref{fig:60045posM3_scans} the allowed regions in the 
$(\theta_{24}, M_0)$ plane for $\tan\beta = 45$, with fixed 
$M_3 = -A_0 = 600$ GeV
and $\mu>0$.
Here we note that the requirement of a neutralino LSP constrains 
the scalar masses $m_0$ to be larger than the minimal values obtained for
moderate $\tan\beta$. In the central continent of 
Figure \ref{fig:60045posM3_scans}, we note a channel of points with 
$\Omegahh < 0.09$; this is the $A^0$-resonance funnel, where dark matter
annihilation is driven chiefly by $s$-channel pseudoscalar exchange, with
$\Omegahh = 0.11\pm 0.02$ regions on either side.
This produces
an island, centered near $\theta_{24}/\pi = 0.05$ and $m_0 = 700$ GeV and 
separate from the main CMSSM-connected continent, which is a confluence 
between the stau co-annihilation, low-$\mu$, and $A^0$ funnel mechanisms 
for efficient dark matter annihilation. However, this island is disfavored
at 95\% level by the constraint from BR$(B_s \rightarrow \mu^+\mu^-) < 1.1 
\times 10^{-8}$. More generally, this constraint eliminates all of the stau co-
annihilation and bulk region models on the central and left continents.
On the right continent, however, a large number of stau co-annihilation models
on the small-$m_0$ edge do survive, with values of $\mu$ ranging up to 700 GeV.
The top edges of the central and right continents are small-$\mu$ regions,
as before, while on the left edge of the right continent, near
$\theta_{24}/\pi = 0.36$, both $s$-channel $A^0$ exchange and a significant 
higgsino content of the LSP play a role in dark matter annihilation.

\begin{figure}[!tbp]
\includegraphics[width=0.7\textwidth]{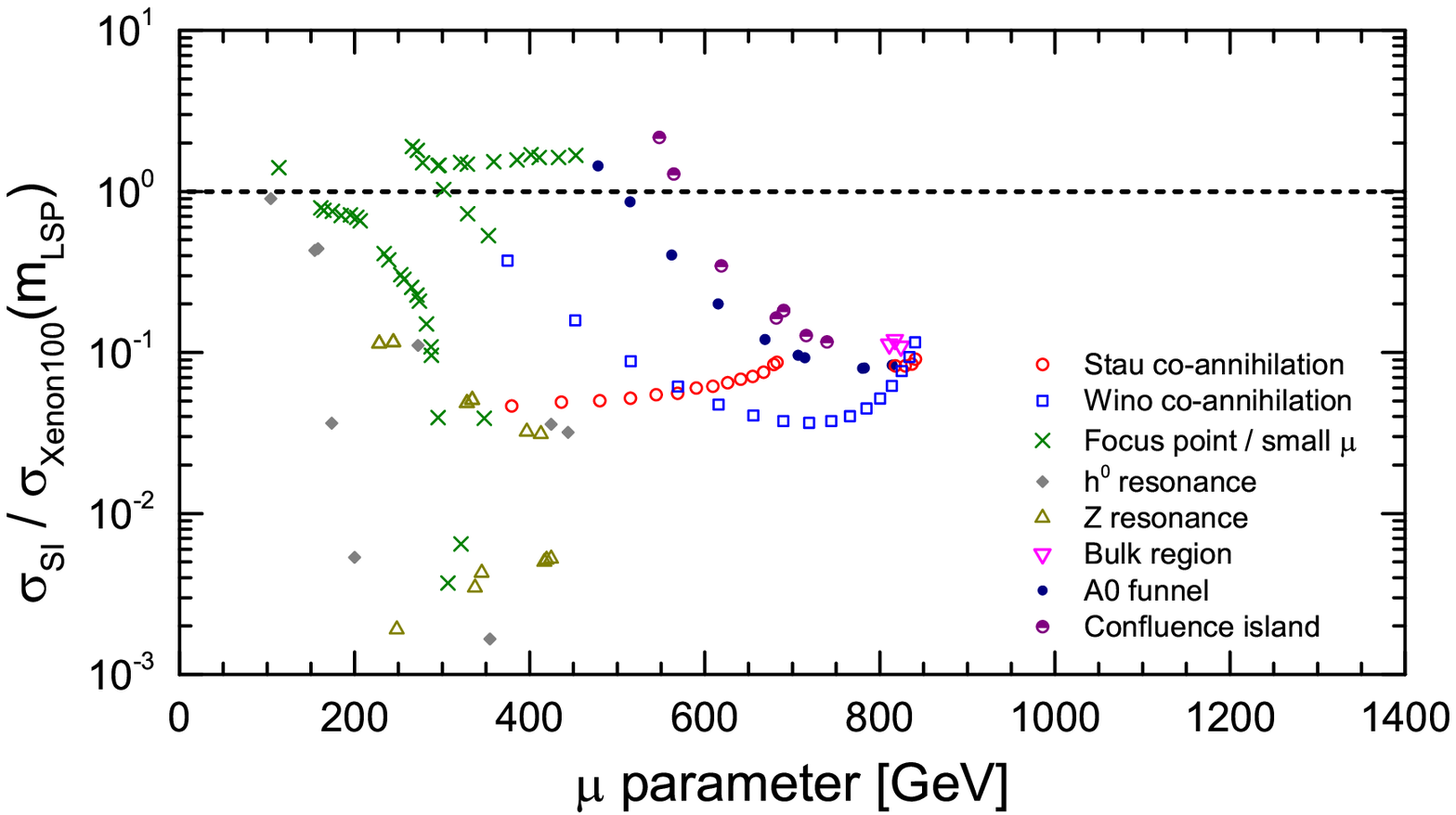} \\
\includegraphics[width=0.7\textwidth]{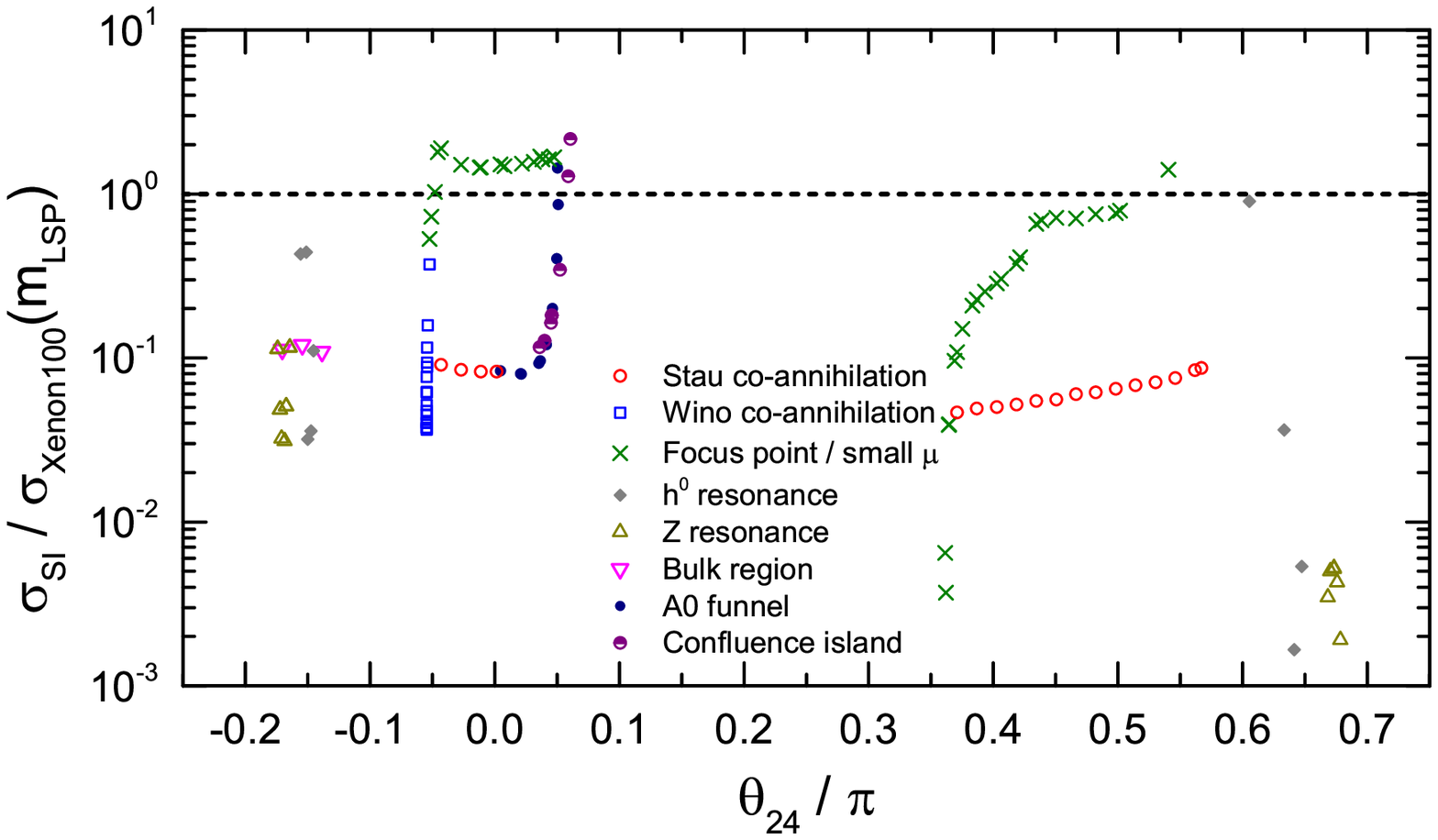}
\caption{\label{fig:60045posM3_SIcross}
The ratio of the spin-independent LSP-nucleon cross section to the 
current XENON100 limit for the model points in Figure 
\ref{fig:60045posM3_scans} (with fixed $M_3=-A_0 =600$ GeV, $\tan \beta = 45$, 
and $\mu>0$ and varying $\theta_{24}, 
m_0$) that have $\Omegahh = 0.11$. Different symbols are 
used for the model points according to which dark matter annihilation 
channels are most important in the early universe.}
\end{figure}

The ratio $\sigma_{SI}/\sigma_{\rm XENON100}$ 
for these models with $\tan\beta=45$
are shown in Figure \ref{fig:60045posM3_SIcross}.
Again we note that models with small $\mu>0$ are most challenged by the 
XENON100 limits on the central, CMSSM-like continent. The models on the 
left side of the right continent, where the pseudoscalar exchange assists 
the higgsino content of the LSP in dark matter annihilations, tend to be 
far below the nominal XENON100 bounds even with the standard 
interpretation of those limits. As before, models with wino 
co-annihilation, $A^0$ resonance, $h^0$ resonance, and $Z$ resonance as 
the dark matter annihilation mechanisms tend to be unchallenged by 
present direct detection limits as well. 
Although we did not encounter models in which the correct
relic abundance of dark matter is brought about by stop or sbottom
co-annihilations \cite{stopcoannihilations,sbottomcoannihilations}, 
there is no general reason why they should not exist,
in particular if the assumption of a common scalar squared mass $m_0^2$
were relaxed.

\clearpage

\section{Explorations with fixed $M_1$}
\setcounter{footnote}{1}
\setcounter{equation}{0}

In this section, we consider alternative slices in parameter space with $M_1$ held fixed. This implies that in most of the parameter space,
the LSP has a mass that does not vary greatly within 
the graphs to be presented below. With fixed $M_1$, the parameterization of
eqs.~(\ref{eq:24parametrizationa})-(\ref{eq:24parametrizationc}) can be 
written as
\beq
M_2 \,=\, M_1 \left( \frac{1 + 3\tan \theta_{24}}{1 + \tan \theta_{24}} 
\right) ,\qquad\quad
M_3 \,=\, M_1 \left( \frac{1 -2 \tan \theta_{24}}{1 + \tan \theta_{24}} 
\right) .
\eeq
Therefore, fixing $M_1$ implies that 
$M_2$ and $M_3$ become very large when 
$\theta_{24}/\pi$ approaches $-1/4$ and $3/4$, so we again choose 
those asymptote values as the boundaries of the range, and again the parameter space splits into three main continents. 
At 
$\theta_{24}/\pi \approx -0.102$, one has $M_2$ approaching 0, so models 
near this vertical line are excluded, 
providing an ocean between the two 
continents. 
Similarly, at $\theta_{24}/\pi \approx 0.148$, one has $M_3$ approaching 
0, so models near this vertical line are likewise excluded, providing 
another gap between the two continents of viable parameter space 
on either side. 
The boundaries of the continents on either side of the $M_3 = 0$ line
are actually determined by the fact that small $|M_3/M_2|$ leads to small 
$|\mu|^2$, 
as can be seen from eqs.~(\ref{eq:m2Z}) and (\ref{eq:m2Hu}).

\begin{figure}[!tbp]
\begin{center}
\includegraphics[width=0.72\textwidth]{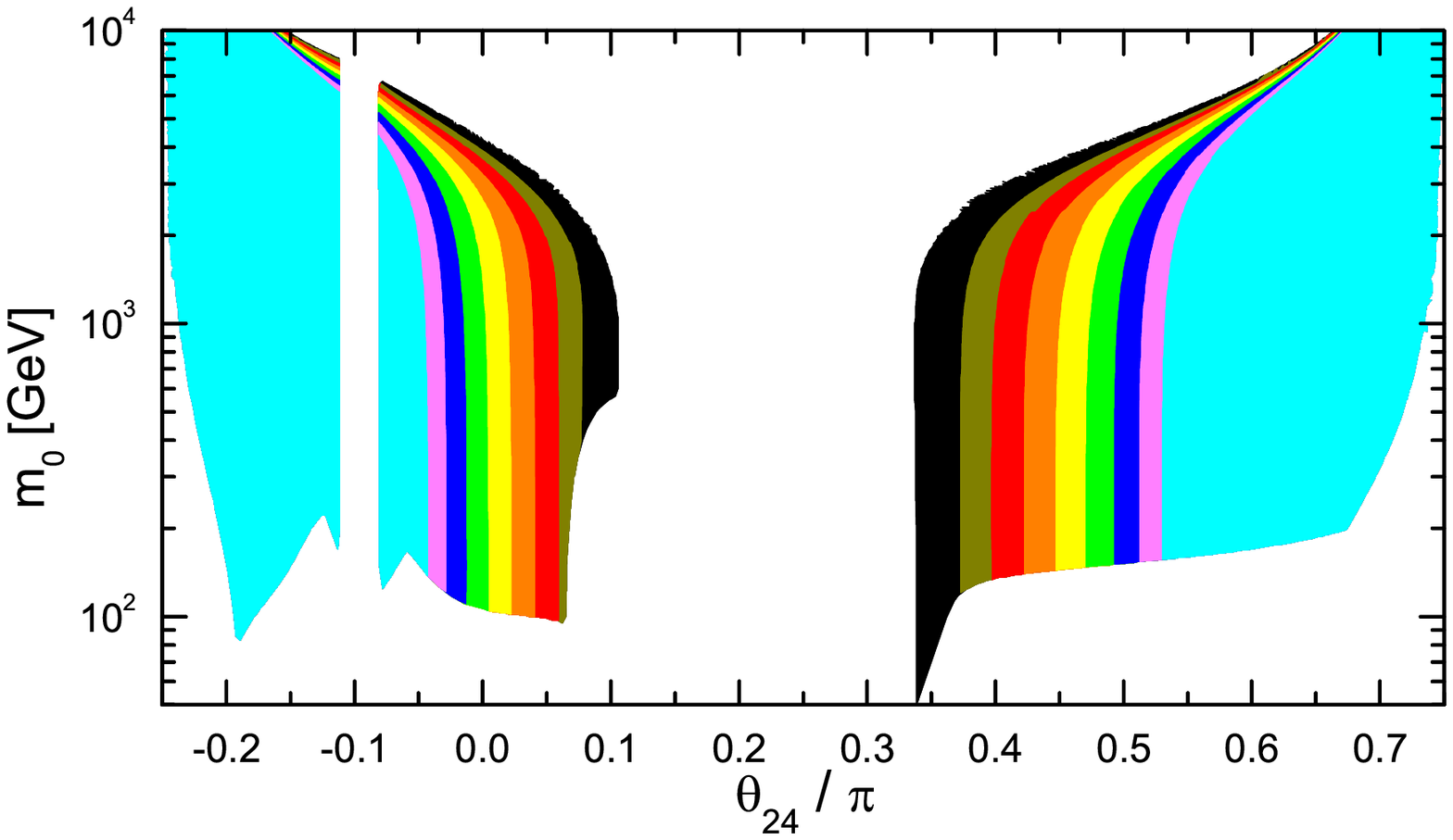}
\\
\includegraphics[width=0.72\textwidth]{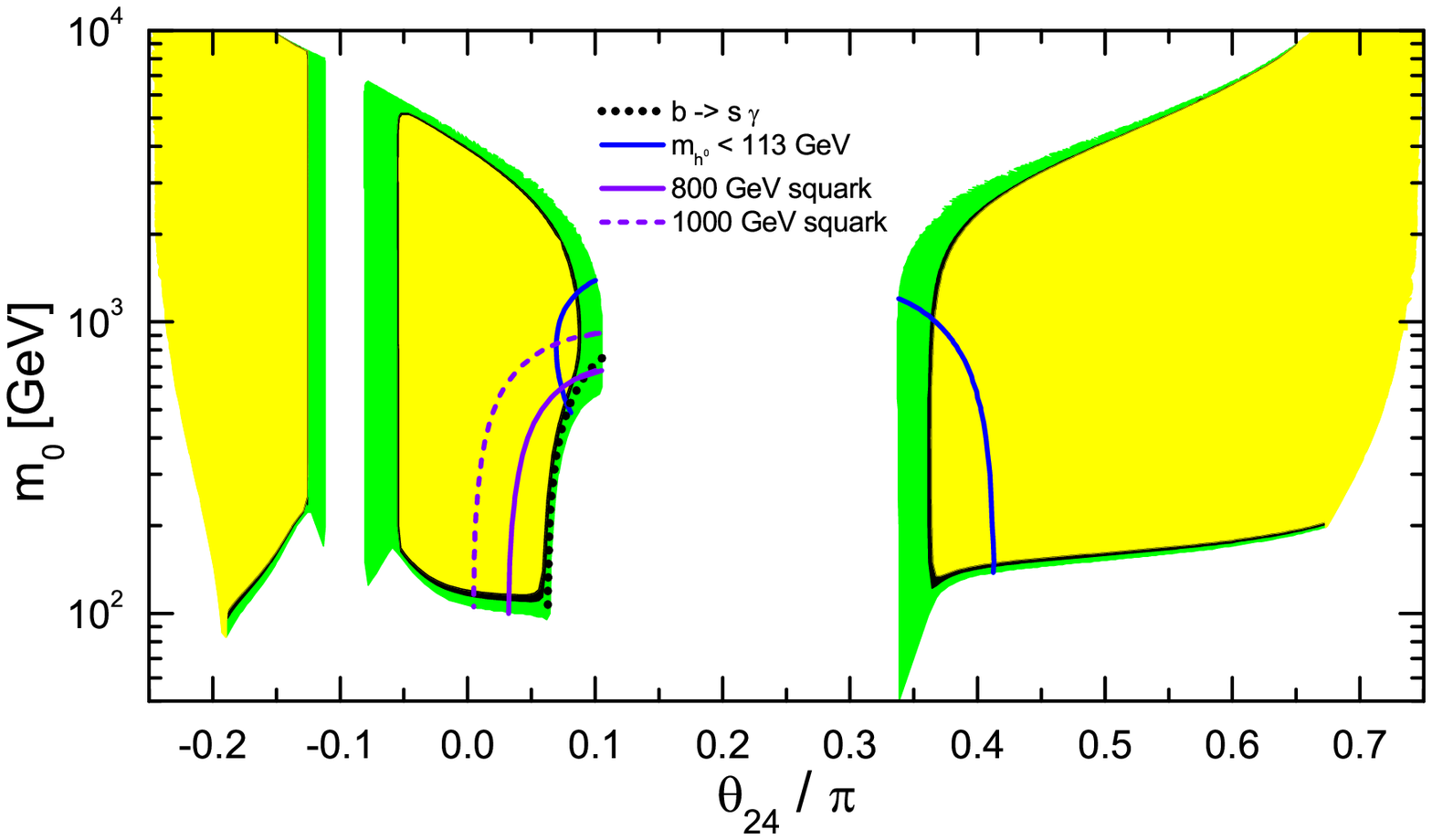}
\end{center}
\caption{\label{fig:50010posM1_scans}
Maps of the $\mu$ parameter (upper plot) and the predicted thermal relic 
abundance of dark matter $\Omegahh$ (lower plot) in the $\theta_{24}, 
m_0$ parameter space with fixed $M_1=-A_0 =500$ GeV, $\tan \beta = 10$, 
and $\mu>0$. In the upper plot, the dark (black) regions closest to the  
center represent $\mu < 300$ GeV, and successive regions away from the 
center correspond to $300$ GeV $ < \mu < 400$ GeV (brown) and $400$ GeV $ 
< \mu < 500$ GeV (red) and so on up to the last region (cyan) which 
corresponds to $\mu > 1000$ GeV. Regions left blank do not have viable 
electroweak symmetry breaking, do not have a neutralino as the LSP, or 
have superpartners that are too light, as described in the text. In the 
lower plot, the thin dark regions (black) correspond to the observed 
range $0.09 < \Omegahh < 0.13$. The large interior regions (yellow) 
correspond to $\Omegahh > 0.13$, while the darker shaded exterior region 
(green) has $\Omegahh < 0.09$. Also indicated are contours for 
$m_{h^0}<113$ GeV, for BR$(b \rightarrow s \gamma)$ from 
eq.~(\ref{eq:bsgamma}), and for $m_{\tilde q} = 800, 1000$ GeV corresponding 
roughly to the present LHC data reach, with regions closer to the center disfavored in each case. 
}
\end{figure}

We consider first models with fixed $M_1 = -A_0 = 500$ GeV, $\tan\beta = 10$,
and $\mu>0$. The maps of $\mu$ and $\Omegahh$ for these models are shown in
Figure \ref{fig:50010posM1_scans}.
The allowed regions are qualitatively similar to the case discussed above for
Figure \ref{fig:60010posM3_scans}, except that in this case there is
a region in which light top squarks mediate the dark matter annihilation,
for $0.055 \lsim \theta_{24}/\pi \lsim 0.08$ and 100 GeV $\lsim m_0 \lsim 600$
GeV, as in ref.~\cite{compressedSUSY}. This region is now likely to be 
ruled out\footnote{However, similar models with non-universal scalar masses, in particular with larger gluino and first- and second-family squark masses, could evade these LHC searches. The 
LHC signatures for this scenario would include, besides the usual jets 
plus $\missingET$ events, events with same-sign top quarks 
\cite{samesigntops} and eventually striking resonant diphoton decays of 
stoponium \cite{stoponium1,stoponium2}, a bound state of top squarks. 
Stoponium is long-lived enough to form bound states in these models 
because of the absence of two-body flavor-preserving decays of the top 
squark.} 
by LHC searches because the gluino is lighter than 660 GeV here, and the average 
squark mass is not much higher, although there is no experimental search 
specifically
sensitive to this particular
version of compressed 
supersymmetry.
More generally, the regions of the plane
that the current LHC published searches are sensitive to are roughly indicated 
by
the $m_{\tilde q} = 800$, 1000 GeV contours shown in the lower plot of Figure
\ref{fig:50010posM1_scans}. Other regions have heavier squarks and/or gluinos.
Note in particular that the left side of the right continent has 
gluino masses above 1.1 TeV, and increasing as one moves to the right.
The region from $0.11 \lsim \theta_{24}/\pi \lsim 0.33$ where one might 
have expected 
to find lighter gluinos was already
excluded because $-m_{H_u}^2$ is negative at the TeV scale, precluding 
electroweak 
symmetry breaking. 
Because the gluino mass increases as one moves away from the 
center of the plots, $\mu$ also increases, and unlike the fixed-$M_3$ plots of 
the previous section, there are now models with $\mu$ well above 1 TeV. This 
illustrates the fine-tuning price of moving to gluino and squark
masses far above the LHC-accessible regions.

Also indicated in Figure \ref{fig:50010posM1_scans} are regions that are 
disfavored by the BR$(b\rightarrow s \gamma)$ and $m_h$ limits. Recall, however,
that the first of these can be evaded by intrinsically supersymmetry-breaking 
flavor violation. Moreover, it is now weaker than the probable direct reach of 
LHC in this particular parameter space slice. 
Note that these constraints eliminate the stau co-annihilation
region with smaller $\mu$ on the CMSSM continent, 
but leave regions with $\mu> 450$ 
GeV on the right continent. 
The focus-point/small-$\mu$ regions on the top edges of the continents are 
untouched by these constraints, and we note that, as before, moving from the 
CMSSM focus point case to $\theta_{24}>0$
reduces the values of $m_0$ required for small $\mu$ and favorable dark matter,
thus arguably substantially decreasing the fine-tuning cost. However, in the 
fixed $M_1 = 500$ GeV parameter space, the $m_h$ constraint still imposes the 
requirement that $m_0$ is well above 1 TeV. Taking larger $M_1$ would ameliorate 
this, as should be clear from the fixed $M_3=600$ GeV cases studied above.

\begin{figure}[!tbp]
\includegraphics[width=0.7\textwidth]{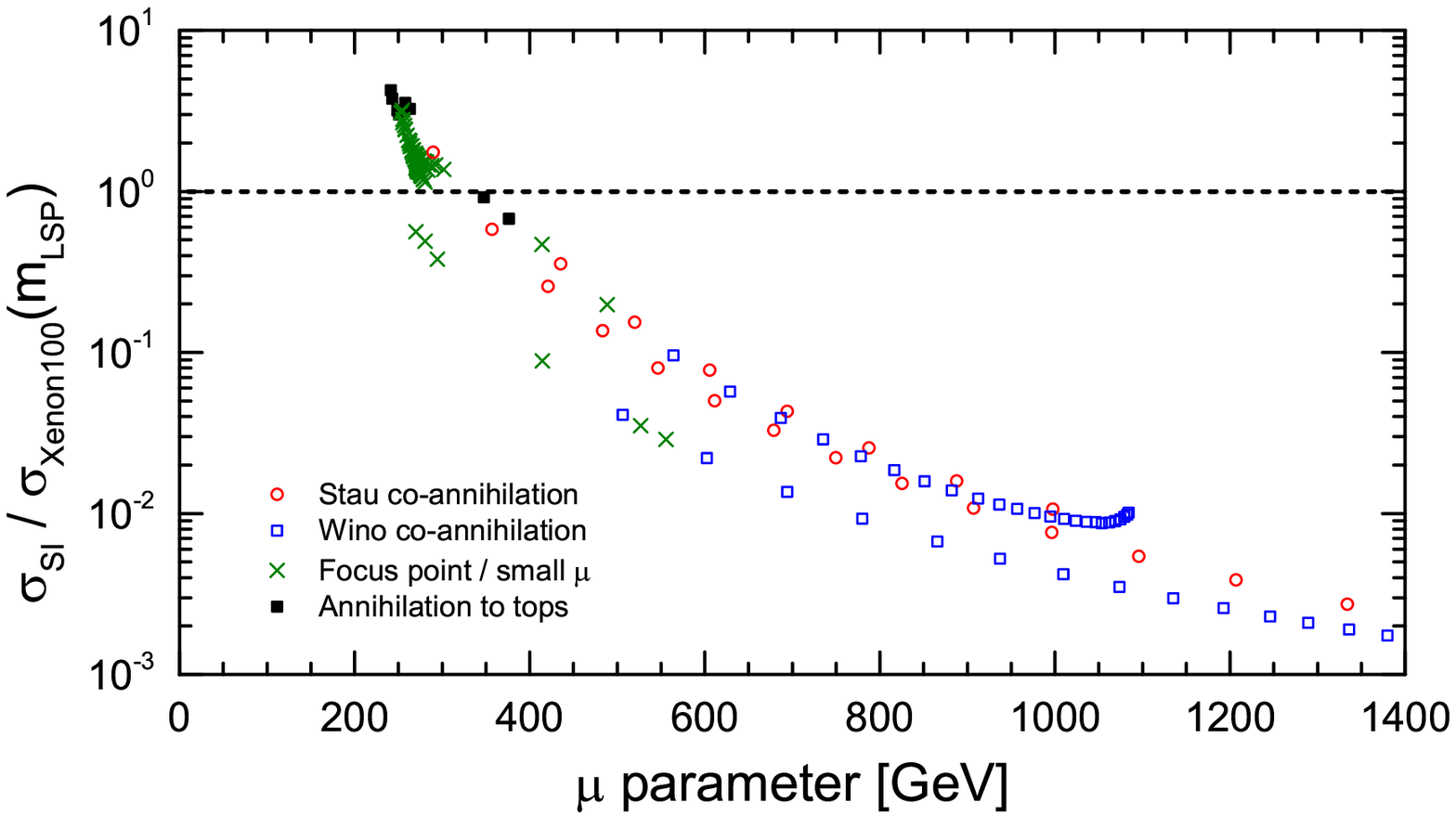} \\
\includegraphics[width=0.7\textwidth]{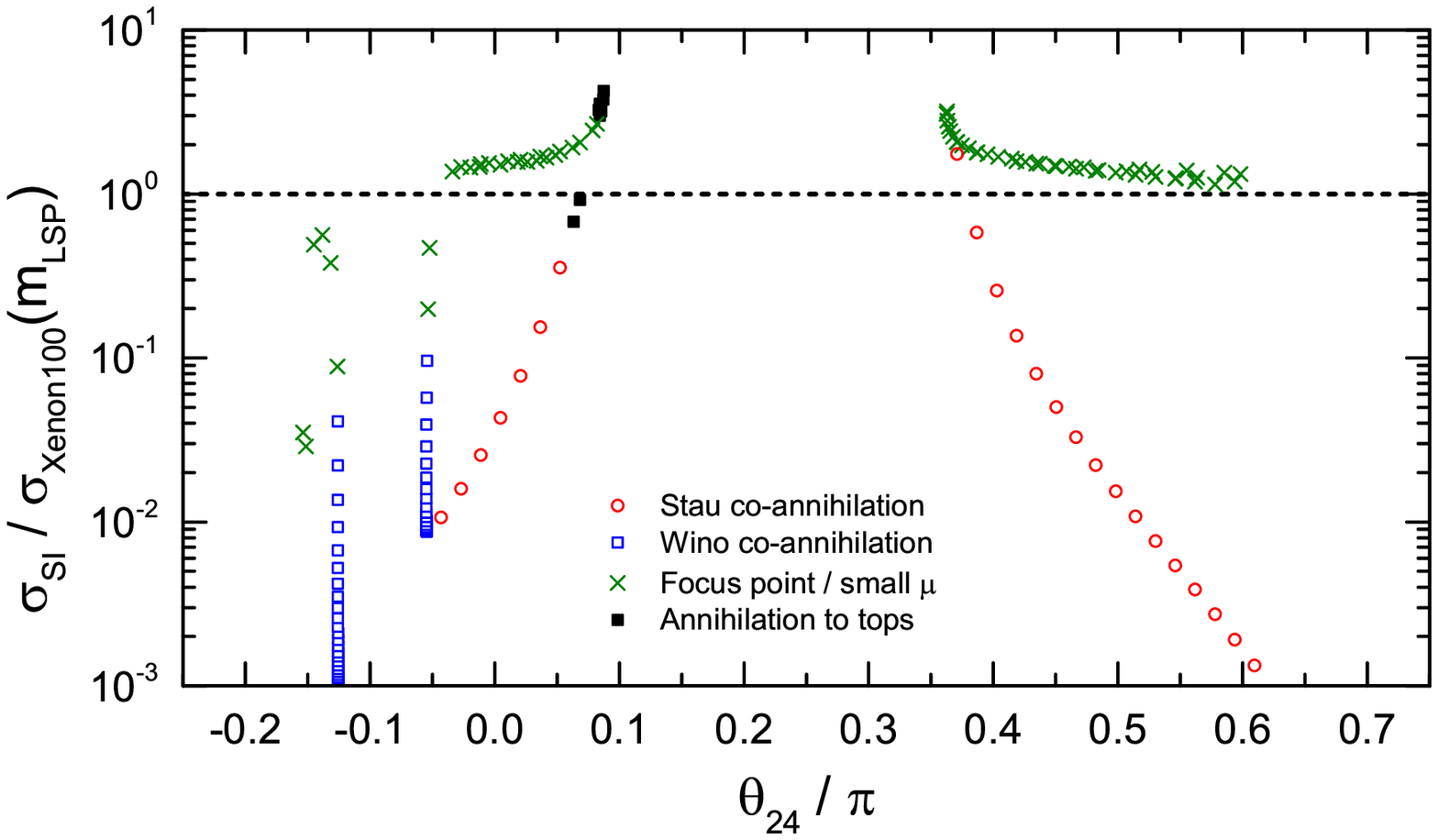}
\caption{\label{fig:50010posM1_SIcross}
The ratio of the spin-independent LSP-nucleon cross section to the 
current XENON100 limit for the model points in Figure 
\ref{fig:50010posM1_scans} (with fixed $M_1=-A_0 =500$ GeV, $\tan \beta = 10$, 
and $\mu>0$ and varying $\theta_{24}, m_0$) that have  
$\Omegahh = 0.11$. Different symbols are 
used for the model points according to which dark matter annihilation 
channels are most important in the early universe.
}
\end{figure}

Figure \ref{fig:50010posM1_SIcross} shows the ratio of the 
spin-independent LSP-nucleon cross section to the 
current XENON100 limit for the model points in Figure 
\ref{fig:50010posM1_scans}.
For fixed $M_1$, we see the well-known fact 
that the correlation between the higgsino content of the LSP, 
inversely proportional to $|\mu/M_1|$ for these models, and the
spin-independent cross-section is quite robust,
with models having $\mu<110$ GeV strongly 
challenged 
within the standard interpretation 
of the LSP as dark matter (but not ruled out, as noted above).
This includes especially the small-$\mu$ models with smaller $m_0$ (i.e.,
the models near $\theta_{24}/\pi = 0.07$ and $0.37$. Conversely, stau 
co-annihilation and wino co-annihilation models not otherwise ruled out
are fine, and models with $\mu > 300$ are two orders of magnitude away from the
sensitivity in direct dark matter experiments needed to challenge them, within 
this slice of parameter space.

\begin{figure}[!tbp]
\includegraphics[width=0.72\textwidth]{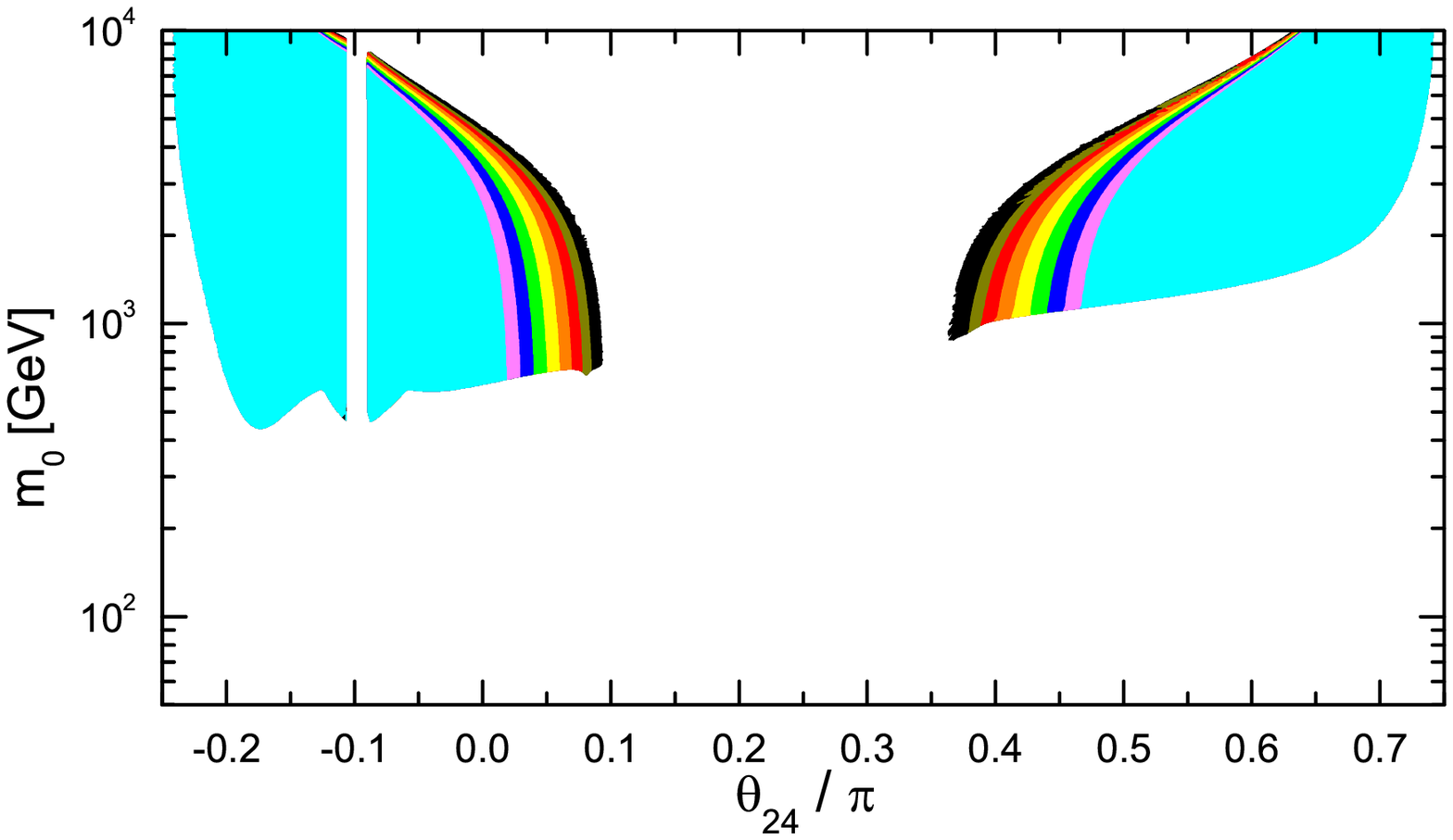} \\
\includegraphics[width=0.72\textwidth]{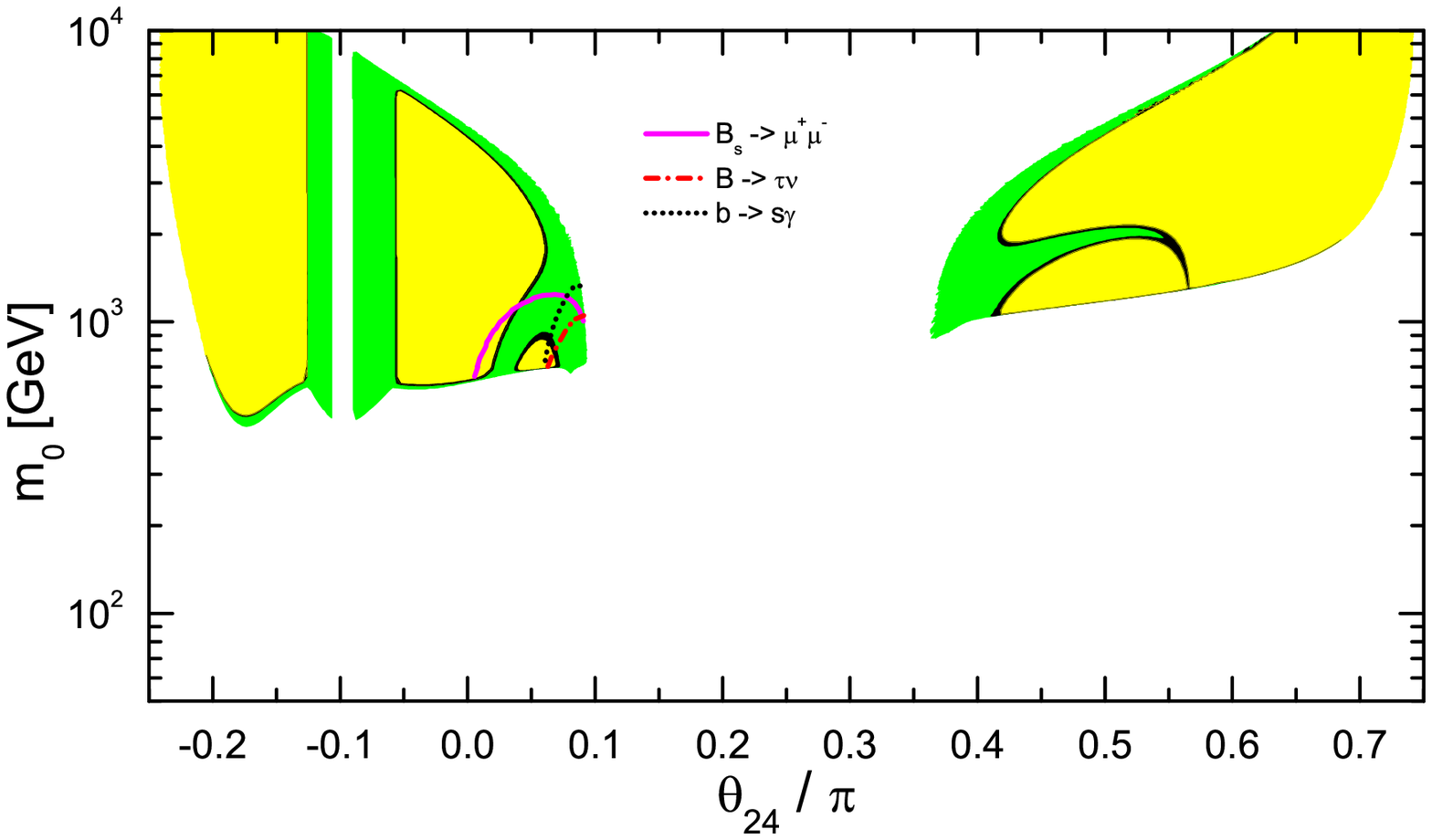}
\caption{\label{fig:90045posM1_scans}
Maps of the $\mu$ parameter (upper plot) and the 
predicted thermal relic abundance of dark matter $\Omegahh$ (lower plot) in the 
$\theta_{24}, m_0$ parameter space with fixed $M_1=-A_0 =900$ GeV, $\tan 
\beta = 45$, and $\mu>0$.
The correspondence of shaded regions to $\mu$ values in the upper plot and to
$\Omegahh$ in the lower plot are the same 
as in Figure \ref{fig:50010posM1_scans}.
Also indicated are contours as described in the text for
the branching ratios of 
$B_s\rightarrow \mu^+\mu^-$
and $B\rightarrow \tau\nu$
and $b \rightarrow s \gamma$. 
}
\end{figure}

\begin{figure}[!tbp]
\includegraphics[width=0.7\textwidth]{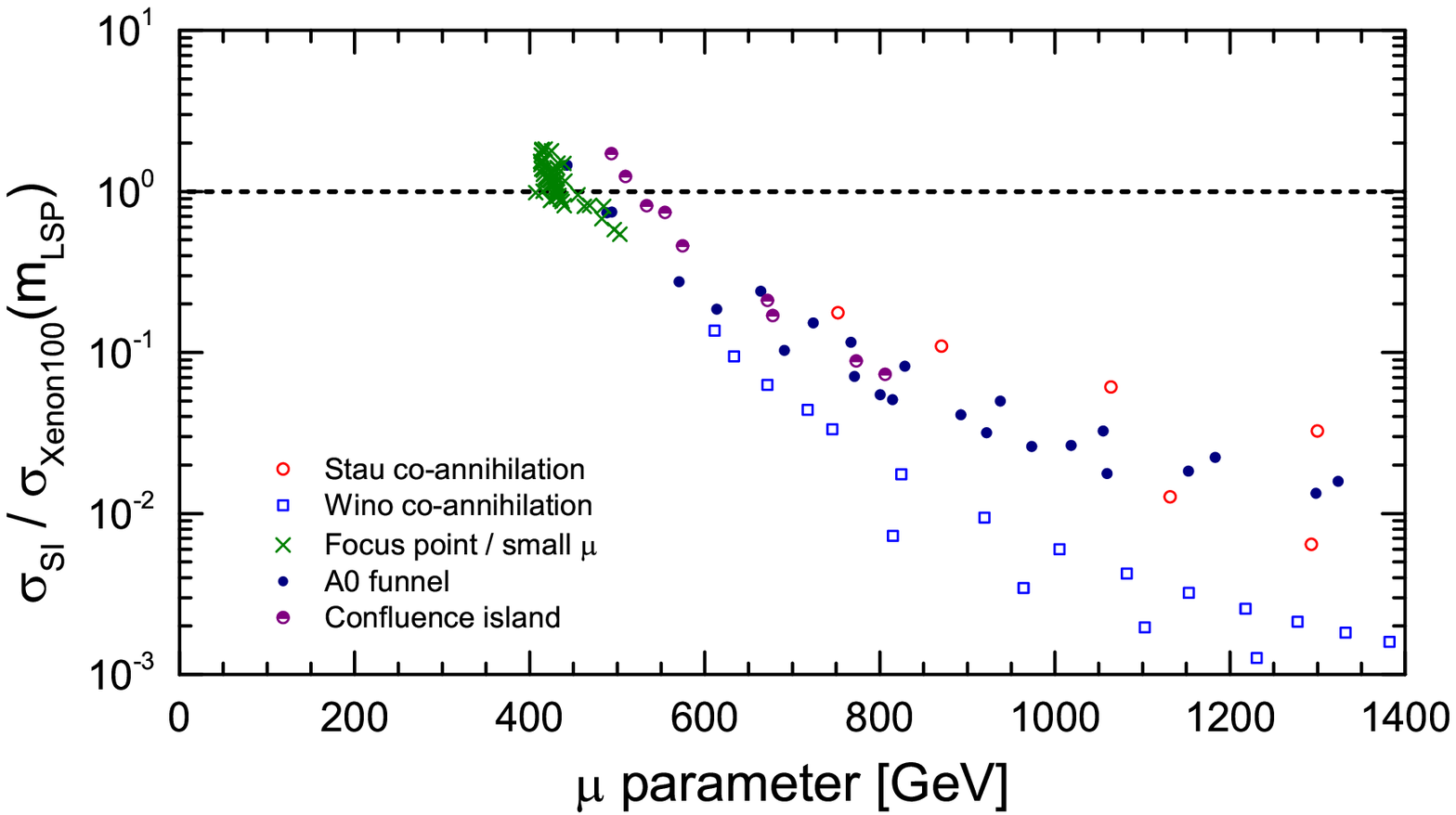} \\
\includegraphics[width=0.7\textwidth]{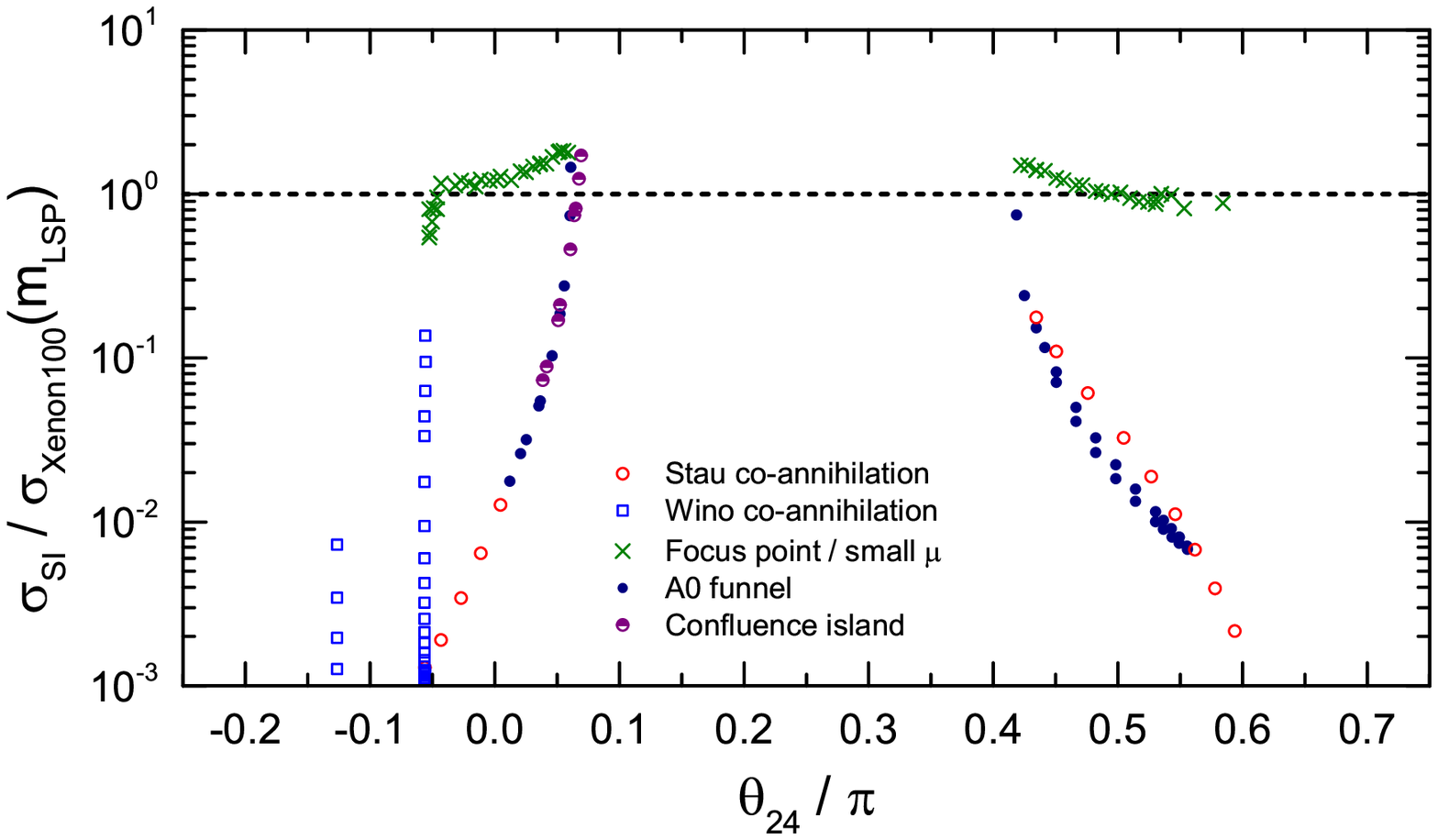}
\caption{\label{fig:90045posM1_SIcross}
The ratio of the spin-independent LSP-nucleon cross section to the 
current XENON100 limit for the model points in Figure 
\ref{fig:90045posM1_scans} (with fixed $M_1=-A_0 =900$ GeV, $\tan \beta = 45$, 
$\mu>0$ and varying $\theta_{24}, m_0$) that have $\Omegahh = 0.11$. 
Different symbols are 
used for the model points according to which dark matter annihilation 
channels are most important in the early universe.}
\end{figure}

As a final exploration, we consider models with fixed $M_1 = -A_0 = 900$ GeV 
and $\tan\beta = 45$, with $\mu>0$. The maps of $\mu$ and $\Omegahh$ are 
shown in Figure \ref{fig:90045posM1_scans} as before.
Notable features include the facts that regions with small 
$\mu$ are much thinner
within this parameter space, and there are no viable regions with $m_0$ less 
than about 500 GeV. As in the case of fixed $M_3$ with 
large $\tan\beta$ in the previous section, there is a confluence island,
on the shores of which $A^0$ resonance, stau co-annihilation, and small-$\mu$
play a role in reducing the dark matter density. However, again the
$B_s\rightarrow \mu^+\mu^-$ constraint strongly disfavors this island,
and the smaller $\mu$ part of it is also disfavored by $B\rightarrow \tau\nu$
and $b \rightarrow s \gamma$. On the right continent, there is 
another large island nearly split off from the main continent by a channel consisting of an $A^0$
resonance funnel region. On the top edge 
of this island, pseudoscalar exchange plays the main role in the annihilation of
dark matter, while on the lower edge it is mainly stau co-annihilation.
The entire $\Omegahh < 0.13$ part of this continent is free from the indirect
constraints eq.~(\ref{eq:Btaunu})-(\ref{eq:mhlimit}). It is also not challenged by dark matter direct detection, except perhaps for the top edge with 
small $\mu$, as can be seen in 
Figure \ref{fig:90045posM1_SIcross}.
This is more generally true of 
all models within this slice of parameter space with
$\mu > 180$ GeV.


\section{Outlook\label{sec:outlook}}
\setcounter{footnote}{1}
\setcounter{equation}{0}

As LHC searches continue to push the bounds on superpartner masses 
higher, there appears to be growing tension between the weak scale $m_Z$ 
and the dimensionful supersymmetry-breaking terms that determine it from 
the effective potential. Non-universal gaugino masses provide a way of 
ameliorating this fine-tuning issue. In this paper we showed that there 
are large regions of parameter space in which $|\mu|$ as determined by 
the supersymmetry-breaking Lagrangian is naturally small, while scalar 
masses can be much less than the several TeV scale found in the 
traditional focus-point region in the MSSM. These models do evade the 
present bounds set by the LHC, but are now being confronted by dark 
matter direct detection experiments. However, exclusion is not possible 
with present XENON100 bounds for most of the cases we looked at, even if 
$R$-parity is conserved and the lightest neutralino is really the dark 
matter. As is well-known, since the spin-independent cross section 
increases with higgsino content of the LSP, which in turn increases as 
$|\mu|$ decreases, future dark matter direct detection searches will 
probe the more natural models first.

In this paper, we also showed how non-universal gaugino masses can 
provide for a bulk region in which $t$-channel slepton exchange (without 
significant co-annihilations) plays the dominant role in annihilating the 
dark matter down to acceptable levels in the early universe. This is 
interesting because this region is severely endangered or perhaps extinct 
within the CMSSM after LHC and LEP bounds are considered. We further 
mapped out the regions in which stau co-annihilations and Higgs resonance 
play the most important role in dark matter annihilations; these regions 
differ in interesting ways from the corresponding CMSSM scenarios, and 
are not always continuously connected to them. In order to keep our study 
finite, we did not consider scalar mass non-universality here, but other 
works (see for example \cite{arXiv:0905.0107} and references therein) 
have showed that non-universal Higgs mass models have large variations in 
the dark matter relic density and spin-independent cross-sections. If 
supersymmetry is discovered at the LHC and in direct dark matter 
detection experiments, it will be a challenge to understand 
to what extent non-universalities in the soft supersymmetry breaking 
parameters play a role in determining the phenomenological collider and 
dark matter properties of the theory.

As this paper was being prepared, the ATLAS and CMS detector 
collaborations announced hints \cite{LHCHiggs} of a Higgs scalar boson 
signal in the vicinity of $M_h = 125$ GeV. If these hints are 
confirmed, the impact on the models considered here would be profound, as 
most of the parameter space cannot accommodate such large $M_h$ values. If 
one assigns a combined
3 GeV uncertainty (note that this is a somewhat arbitrary estimate,
including both the theoretical uncertainty of about 2 GeV
\cite{Mhuncertainty} and the quite preliminary LHC uncertainty)
to $M_h$, and therefore considers regions with predicted
$M_h \gsim 122$ GeV to be acceptable, 
then among the models we considered only those with 
moderately 
negative $\theta_{24}$ (on the main continent) and large $m_0$ are 
viable. From the point of view of the putative $M_h=125$ GeV signal,
the variation on the usual CMSSM focus point region with negative
$\theta_{24}$ is therefore preferred.
There are several other ways to increase the prediction for 
$M_h$, 
including by raising all of the soft masses (by raising $M_3$ in our 
framework) and by increasing $\tan\beta$. However, among the models considered here, the heavy scalar 
solutions with negative $\theta_{24}$
would seem to be the easiest to make consistent with $M_h \approx 125$ 
GeV. Other mechanisms, such as adding new vector-like supermultiplets 
\cite{vectorlike1}-\cite{vectorlike5}, would affect our results in 
qualitative ways.

\bigskip \noindent
{\it Acknowledgments:}
This work was supported in part by the National Science Foundation grant 
numbers PHY-0757325 and PHY-1068369. The work of JEY was supported in 
part by the 
U.S. 
National Science Foundation, grant NSF-PHY-0705682, the LHC Theory Initiative, 
Jonathan Bagger, PI.


\end{document}